\documentclass[a4paper,fleqn,times]{cas-sc}
\usepackage{etex}
\usepackage{ifthen}
\usepackage{xifthen}
\usepackage{xkeyval}
\usepackage{xstring}
\usepackage{etoolbox}
\usepackage{array}
\usepackage{amsthm}
\usepackage{amsbsy}
\usepackage{amsmath}
\usepackage{mathtools}
\usepackage{amsfonts}
\usepackage{amssymb}
\usepackage{bm}
\usepackage{xcolor}
\usepackage{graphicx}
\usepackage{relsize}
\usepackage{stmaryrd}                                
\RequirePackage{tabto}
\RequirePackage{environ}
\RequirePackage{paralist}
\RequirePackage{enumitem}
\RequirePackage{etaremune}
\RequirePackage{listings}
\RequirePackage{mdwlist}
\RequirePackage{verbatim}
\RequirePackage{fancyvrb}
\RequirePackage[noEnd=false,indLines=false,italicComments=false,commentColor=black]{algpseudocodex}
\usepackage{dcolumn}
\usepackage{booktabs}
\usepackage{multicol}
\usepackage{multirow}
\usepackage{rccol}
\usepackage{tabularx}
\usepackage{suffix}
\usepackage{lipsum}
\usepackage{siunitx}
\usepackage{comment}
\usepackage{csquotes}
\usepackage{enumitem}
\usepackage{datetime}
\usepackage[noabbrev]{cleveref}
\usepackage{fix-cm}
\usepackage{xspace}
\usepackage{epigraph}
\usepackage{lastpage}
\usepackage{adjustbox}
  \usepackage[utf8]{inputenc}
  \usepackage[T1]{fontenc}

\usepackage{tcolorbox}
\usepackage[sort&compress,longnamesfirst,numbers]{natbib}
\tcbuselibrary{breakable}
\DeclareFontFamily{U}{dutchcal}{\skewchar\font=45 }
\DeclareFontShape{U}{dutchcal}{m}{n}{<-> s*[1.0] dutchcal-r}{}
\DeclareFontShape{U}{dutchcal}{b}{n}{<-> s*[1.0] dutchcal-b}{}
\DeclareMathAlphabet{\mathlcal}{U}{dutchcal}{m}{n}
\SetMathAlphabet{\mathlcal}{bold}{U}{dutchcal}{b}{n}
\newcommand{\vectorsym}[1]{\ensuremath{\boldsymbol{#1}}}
\newcommand{\tensorsym}[1]{\ensuremath{\boldsymbol{#1}}}
\newcommand{\matrixsym}[1]{\ensuremath{\boldsymbol{#1}}}
\newcommand{\average}[1]{\ensuremath{\left\{\!\!\left\{#1\right\}\!\!\right\}}}
\newcommand{\jump}[1]{\ensuremath{\left\llbracket#1\right\rrbracket}}

\newcommand{\diff}[1]{\,\mathrm{d}#1}
\newcommand{\dn}[2]{\dfrac{\diff#1}{\diff#2}}

\newcommand{\dpn}[2]{\dfrac{\partial#1}{\partial#2}}

\newcolumntype{C}{>{\centering\arraybackslash}X}
\newcolumntype{N}[2]{>{\centering\arraybackslash}R[.][.]{#1}{#2}}
\def\figurespath{./}
\definecolor{base}{rgb}{0.25,0.50,0.75}
\ExplSyntaxOn
\keys_set:nn { stm / mktitle } { nologo }
\ExplSyntaxOff
\makeatletter
\newtoks\arxiv@@algname@current
\def\createalglabel#1#2{%
  \arxiv@@algname@current={#2}%
  \edef\arxiv@temp{\the\arxiv@@algname@current}%
  \@bsphack\protected@write{\@auxout}{}{\string\arxiv@@newalglabel{#1}{\expandafter\strip@prefix\meaning\arxiv@temp}}\@esphack%
}
\def\arxiv@@newalglabel#1#2{%
  \expandafter\xdef\csname arxiv@@alglabel@#1\endcsname{#2}%
}
\algnewcommand\algorithmiccall{\textbf{call}}
\algnewcommand\algorithmicnot{\textbf{not}}
\algnewcommand\algorithmicand{\textbf{and}}
\algnewcommand\algorithmicnil{\textbf{nil}}
\algnewcommand\algorithmicor{\textbf{or}}
\algnewcommand\algorithmicto{\textbf{to}}
\algnewcommand\algorithmicswitch{\textbf{switch}}
\algnewcommand\algorithmiccase{\textbf{case}}
\algnewcommand\algorithmicuse{\textbf{use}}
\algnewcommand\algorithmicinput{\textbf{input}}
\algnewcommand\UseAlgorithm[1]{\algorithmicuse~Algorithm~\ref{#1}:~\textrm{\csname arxiv@@alglabel@#1\endcsname}}
\algnewcommand\UseVariable[2]{\algorithmicuse~\textit{#1} (\textrm{#2})}

\algnewcommand\CiteAlgorithm[1]{Algorithm~\ref{#1}}
\algnewcommand\InputVariable[2]{\algorithmicinput:~\textit{#1}}
\algnewcommand\OutputVariable[2]{\algorithmicoutput~\textit{#1}}
\algnewcommand\ReturnVariable[2]{\algorithmicreturn~\textit{#1}}

\algnewcommand\DefineVariable[2]{\textbf{#1} \textit{#2}}
\algnewcommand\CallProcedure[2]{\algorithmiccall~\Call{#1}{#2}}
\algnewcommand\CallFunction[2]{\Call{#1}{#2}}
\algnewcommand\Variable[1]{\textit{#1}}
\algrenewcommand\textproc{\textrm}
\algdef{SE}[SWITCH]{Switch}{EndSwitch}[1]{%
  \algpx@startCodeCommand\algpx@startIndent\algorithmicswitch\ #1\ \algorithmicdo%
}{%
  \algpx@startEndBlockCommand\algpx@endIndent\algorithmicend\ \algorithmicswitch%
}%
\algdef{SE}[CASE]{Case}{EndCase}[1]{%
  \algpx@startCodeCommand\algpx@startIndent\algorithmiccase\ #1%
}{%
  \algpx@startEndBlockCommand\algpx@endIndent\algorithmicend\ \algorithmiccase%
}%
\ifbool{algpx@noEnd}{%
  \algtext*{EndSwitch}%
  \algtext*{EndCase}%
  \pretocmd{\EndSwitch}{\algpx@endIndent}{}{}
  \pretocmd{\EndCase}{\algpx@endIndent}{}{}
}{%
  \algtext*{EndCase}%
  \pretocmd{\EndCase}{\algpx@endIndent}{}{}
}%

\pretocmd{\Switch}{\algpx@endCodeCommand}{}{}
\pretocmd{\Case}{\algpx@endCodeCommand}{}{}

\ifbool{algpx@noEnd}{%
  \pretocmd{\EndSwitch}{\algpx@endCodeCommand[1]}{}{}%
  \pretocmd{\EndCase}{\algpx@endCodeCommand[1]}{}{}%
}{%
  \pretocmd{\EndSwitch}{\algpx@endCodeCommand[0]}{}{}%
  \pretocmd{\EndCase}{\algpx@endCodeCommand[0]}{}{}%
}%
\algsetblock{Procedure}{EndProcedure}{}{1.0em}
\algrenewtext{For}[3]{$\algorithmicfor\ #1 = #2\ \algorithmicto\ #3\ \algorithmicdo$}
\makeatother
\tcbset{%
  my-remarks/.style={%
    coltitle=red,
    colbacktitle=gray!20,
    colupper=red,
    colback=gray!10,
    colframe=base,
    halign title=flush center,
    fonttitle=\large\bfseries\rmfamily,
    arc=1.00pt,
    toptitle=0.20cm,
    bottomtitle=0.20cm,
    lefttitle=0.20cm,
    righttitle=0.20cm,
    top=0.25cm,
    bottom=0.25cm,
    left=0.25cm,
    right=0.25cm,
    boxsep=0.00cm,
    boxrule=2.00pt,
    subtitle style={%
      halign=center,
      colupper=base,
      colback=gray!10,
      boxrule=0.75pt,
      boxsep=0.00cm,
      },
    breakable
  }
}
\tcbset{%
  my-algorithm/.style={%
    coltitle=black,
    colbacktitle=white,
    colupper=black,
    colback=white,
    colframe=black,
    fonttitle=\normalsize,
    fontupper=\small,
    fontlower=\small,
    arc=0.00pt,
    toptitle=0.20cm,
    bottomtitle=0.20cm,
    lefttitle=0.20cm,
    righttitle=0.20cm,
    top=0.25cm,
    bottom=0.25cm,
    left=0.25cm,
    right=0.25cm,
    boxsep=0.00cm,
    boxrule=0.75pt,
    subtitle style={%
      halign=center,
      colupper=base,
      colback=gray!10,
      boxrule=0.75pt,
      boxsep=0.00cm,
      },
    breakable
  }
}
\newcounter{algorithm}
\newenvironment{algorithm}[3]%
  {%
  \refstepcounter{algorithm}%
  \begin{tcolorbox}[%
    my-algorithm,%
    title={%
      \textbf{Algorithm~\thealgorithm:}~\textrm{#2}%
      \ifthenelse{\equal{#1}{}}{\relax}{\label{#1}\createalglabel{#1}{#2}}%
      }%
    ]%
  }%
  {%
  \end{tcolorbox}
}
  {%
  \begin{tcolorbox}[%
    my-remarks,%
    title={PENDING},%
    ]%
  }%
  {%
  \end{tcolorbox}%
}
\begin{document}
%
\def\floatpagepagefraction{1}
\def\textpagefraction{0.001}
\shorttitle{A geometry-adaptive mesh refinement framework for immersed boundary methods}
\shortauthors{Núñez}
\title[mode=title]{A geometry-adaptive mesh refinement framework for immersed boundary methods on hybrid unstructured meshes}
\author[1,2]{Jonatan {Núñez-de la Rosa}}[%
  auid=001,%
  bioid=001,%
  orcid=0000-0002-4541-2768]
\cormark[1]
\ead{jonatan.nunez@upm.es}
\credit{Conceptualization of this study, Methodology, Software}
\author[1,2]{Esteban Ferrer}[%
  auid=002,%
  bioid=002,%
  orcid=0000-0003-1519-0444]
\ead{estaban.ferrer@upm.es}
\credit{Conceptualization, Methodology, Project administration, Funding acquisition}
\author[1,2]{Eusebio Valero}[%
  auid=003,%
  bioid=003,%
  orcid=0000-0002-1627-6883]
\ead{eusebio.valero@upm.es}
\credit{Conceptualization, Methodology, Project administration, Funding acquisition}
\affiliation[1]{%
  organization={School of Aeronautics, Universidad Politécnica de Madrid},
  addressline={Plaza del Cardenal Cisneros 3},
  city={Madrid},
  postcode={28040},
  country={Spain}%
}
\affiliation[2]{%
  organization={Center for Computational Simulation, Universidad Politécnica de Madrid},
  addressline={Campus Montegancedo},
  city={Boadilla del Monte},
  postcode={28660},
  country={Spain}%
}
\cortext[1]{Corresponding author}
%
\begin{abstract}
  In this work, we describe a new preprocessing tool for geometry-adaptive mesh refinement on hybrid unstructured meshes with a target application on immersed boundary methods. The tool has as input an unstructured, hybrid, and conforming mesh generated by an external mesh generation software, and the main goal is to refine this mesh around immersed geometries in such a way that the CFD solver using the immersed boundary method can simulate flow problems in an accurate and efficient manner. The input background mesh can be made of different types of elements, like tetrahedra, hexahedra, prisms, and pyramids, which, unlike Cartesian meshes, permit for a more flexible mesh. Hybrid unstructured meshes enable one to use the immersed boundary technology in a new class of flow problems where the full geometry is decomposed into a fixed geometry part and a changing geometry part. A body-fitted mesh is generated for the fixed geometry while for the changing one is used the immersed boundary method. We simulate several flow problems to test the new meshes, including subsonic flow past a cylinder and subsonic flow past an NACA0012 airfoil, both using finite volume and discontinuous Galerkin methods and solving the Navier--Stokes equations. As an industrial example of our mesh generation, we consider the simulation of a multi-element airfoil: in this case, a mesh generation software generates an unstructured conforming background mesh for the slat and main airfoil, while the flap is placed as immersed geometry in this body-fitted mesh. As accurate and efficient results are sought, this mesh is refined around the flap and then the subsonic flow at high-lift flow conditions is simulated with a finite volume method coupled with an immersed boundary method and using the Reynolds--averaged Navier--Stokes equations. The reported numerical simulations are in good agreement with their corresponding full body-fitted simulations and with experimental data. As a final application example, we study the optimization of the flap position in a two-element airfoil with respect to the lift coefficient. To this end, we have developed an optimization methodology using the open-source GEMSEO optimization suite and the CODA CFD solver.%
  \newline\newline\medskip
  {%
    \scriptsize%
    \textbf{PROGRAM SUMMARY} \\
    {\em Program Title:~}%
      {Adaptive Mesh Refinement for Immersed Boundary Methods} \\
    {\em CPC Library link to program files:~}%
      {(to be added by Technical Editor)} \\
    {\em Developer's repository link:~}%
      {\url{https://github.com/jbnunezd/mesh-refinement-for-ibm}} \\
    {\em Licensing provisions:~}%
      {GPLv3} \\
    {\em Programming language:~}%
      {Fortran 2008} \\
    {\em External routines/libraries:~}%
      {HDF5} \\
    {\em Nature of problem:~}%
      {{AMR-for-IBM} is a new preprocessing tool for geometry-adaptive mesh refinement on hybrid unstructured meshes with a target application on immersed boundary methods. The tool has as input an unstructured, hybrid, and conforming mesh generated by an external mesh generation software, and the main goal is to refine this mesh around immersed geometries in such a way that the CFD solver using the immersed boundary method can simulate flow problems in an accurate and efficient manner. The input background mesh can be made of different types of elements, such as tetrahedra, hexahedra, prisms, and pyramids, which, unlike Cartesian meshes, permit a more flexible mesh.} \\
    {\em Solution method:~}%
      {{AMR-for-IBM} is mainly constructed based on a dynamical linked list that stores the elements of the mesh, allowing full flexibility in the mesh generation. This data structure includes methods for creating, adding, and removing elements, nodes, and faces in the list itself. Hash tables are used mainly to construct mappings between nodes and faces, which allow to find a face if the nodes that constitute this face are provided. A $k$-d tree is used to find the nearest neighbors of the immersed geometry facets to the mesh elements, so the overlap of the geometry facets with the mesh elements can be efficiently computed.} \\
  }
\end{abstract}
%
%
%
\begin{keywords}
  mesh refinement \sep
  hybrid meshes \sep
  immersed boundaries \sep
  computational fluid dynamics
\end{keywords}
%
\maketitle
%
\section{Introduction}\label{sec:introduction}
Computational fluid dynamics has become a powerful tool to simulate fluid flow problems in science and engineering, such as blood flow through the circulatory system~\cite{marom2015a}, airflow around aircrafts or wind turbines~\cite{bazilevs2014a,sclafani2014b,volpiani2024a}, relativistic flows in astrophysics~\cite{nunez2016a,nunez2016b,nunez2018a,berta2024a}, or even plasmas in fusion reactors~\cite{blokland2007a,schneider2015a}. In recent years, numerical schemes have been developed to increase the fidelity of CFD simulations. Whether they are based on finite difference, finite volume, or discontinuous Galerkin methods, CFD solvers have a common feature: they require a computational mesh. To obtain a good solution of the fluid dynamics governing equations, accurate numerical schemes along with high-quality computational meshes are important in the simulation workflow. In industrial applications, complex geometries are ubiquitous, and therefore their good representation is essential. The generation of an appropriate body-fitted mesh for a given simulation problem in fluid dynamics can be tedious and very expensive, requiring not only an excellent mesh generation software but also a skilled simulation engineer.\par
Immersed boundary methods have arisen as an alternative to numerical schemes using body-fitted meshes. These numerical techniques are used in computational fluid dynamics to model the interaction between fluids and rigid or flexible bodies without requiring complex body-fitted mesh generation. Instead of a conformal mesh, a Cartesian background mesh is used and the body geometry is immersed in it. This means that complex structures, such as solid objects or porous media, are represented as part of the fluid computational domain without the need for a grid that conforms to their complicated shapes~\cite{verzicco2023a}. The underlying numerical scheme, whether this is a finite-difference or a finite-volume scheme, or even a discontinuous Galerkin method, suffers little change in case it is combined with an immersed boundary method. In fact, the boundary conditions have to be adapted/modified in order to take into account the immersed body.\par
Immersed boundary methods are classified according to the way in which the interaction between the fluid and the immersed body is carried out. Among the available approaches, the most relevant are the direct forcing approach~\cite{goldstein1993a,fadlun2000a,luo2012a,boukharfane2018a}, the cut-cell methods~\cite{udaykumar2001a,orley2015a}, the ghost cell methods~\cite{tseng2003a,hu2013c}, the sharp interface methods~\cite{ghias2007a,de-vanna2020a}, and the volume penalization schemes~\cite{liu2007a,engels2015a,abalakin2016a}. The latter family of immersed boundary methods has become a promising technique for avoiding the complications of generating body-fitted meshes for intricate geometries and also for simulating moving geometries. In the works of~\citet{mittal2005a,sotiropoulos2014a,griffith2020a,verzicco2023a} more details on the different families of immersed boundary methods can be found. For the fluid flow problems we discuss in this work, the use of these techniques to simulate turbulent aerodynamic flows is discussed in~\cite{capizzano2011a,capizzano2016a,capizzano2018a,peron2021a,constant2024a,nunez2025a,nunez2026a,nunez2026b}.\par
Simulations of aerodynamics flow problems based on immersed boundary methods require a huge number of nodes or elements to accurately predict the fluid flow properties, especially in the local areas close to the wall boundary. Uniform Cartesian meshes are prohibitive in this case because the mesh will need billion nodes/elements to satisfy the requirement previously mentioned, therefore adaptive mesh refinement algorithms are necessary to overcome this issue. Recently, new strategies have been developed to construct adaptive Cartesian meshes for complex three-dimensional geometries in the context of finite difference methods~\cite{wang2020a,luo2025a}, finite volume~\cite{capizzano2018a,peron2021a,meng2022a,constant2024a,nunez2025a}, and discontinuous Galerkin~\cite{ouyang2024a,qi2024a,nunez2025a}.\par
In this work, we have developed a new mesh refinement preprocessing tool that works on hybrid unstructured meshes for immersed boundary methods. The idea is to import a mesh generated by an external mesh generation software. This external mesh is an unstructured, hybrid and conforming mesh, made basically of tetrahedra, hexahedra, prisms, and pyramids. The input mesh is created in such a way that it is body-fitted for certain parts of the geometry, but it is an immersed boundary mesh in locations where other geometries can be placed. Then this mesh is refined only in the locations where the immersed body is located, allowing, unlike Cartesian meshes, more flexibility. In this way, immersed boundary methods can be used in a new family of flow problems where the whole geometry is broken up into a fixed geometry and a changing geometry. A body-fitted mesh is generated for the fixed geometry while for the changing one is placed as an immersed body. This kind of body-fitted/immersed-boundary meshes is of particular interest in industrial aerodynamics.\par
The finite volume and discontinuous Galerkin methods coupled with the immersed boundary volume penalization schemes are used in order to simulate fluid flow problems with complex geometries. To solve the Navier--Stokes and Reynolds--Averaged Navier--Stokes equations, we have employed the CODA CFD solver~\cite{kroll2016a} with the immersed boundary penalization plugin already implemented in a previous work~\cite{nunez2025a}. CODA is a numerical framework for multidisciplinary analysis and optimization of aircraft and helicopters based on advanced and accurate numerical techniques. The solver is being developed as part of a collaboration between the French National Aerospace Research Center (ONERA), the German Aerospace Center (DLR), and Airbus. The spatial discretization schemes used in CODA to solve partial differential equations are the finite volume method, the modal discontinuous Galerkin method, and the discontinuous Galerkin spectral element method. In CODA are also available explicit, implicit, and implicit-explicit time integrators.\par
This work is organized as follows: in \cref{sec:governing-equations} are summarized the governing equations that have been used to simulate fluid flow problems within the CODA CFD solver, namely the Navier--Stokes equations and the Reynolds-averaged Navier--Stokes equations. In \cref{sec:numerical-methods} are briefly discussed the numerical schemes that have been employed to solve the governing equations: for the spatial discretization, these schemes are the finite volume and discontinuous Galerkin methods, and for the time discretization, the backward Euler method; the immersed boundary volume penalization are also introduced. In \cref{sec:mesh-refinement} are explained in detail the most important algorithms that have been developed and implemented in the mesh refinement preprocessing tool. Next, in \cref{sec:numerical-computations} we present several numerical computations performed with the CODA CFD solver coupled with the immersed boundary volume penalization method and using the meshes refined with our preprocessing tool: the subsonic flow past a cylinder, the subsonic flow past an NACA0012 airfoil, the subsonic flow past an MDA30P30N multi-element airfoil, and the optimization of the flap position in a two-element airfoil with respect to the lift coefficient. Finally, in \cref{sec:conclusions}, a summary of this work is presented and the advantages and disadvantages of the immersed boundary volume penalization methods for aerodynamic simulations in industrial scenarios are discussed.
%
\section{Governing equations}\label{sec:governing-equations}
In this work, we have considered the Navier--Stokes and Reynolds--Averaged Navier-Stokes equations. The final goal is to simulate fluid flow problems in industrial aerodynamics with the immersed boundary methods on refined unstructured meshes. We formulate these equations as systems of conservation laws, which is the natural form used by finite volume and discontinuous Galerkin methods.
\subsection{Navier--Stokes equations}\label{sec:governing-equations:navier-stokes}
The Navier--Stokes equations are a system of partial differential equations that describes the motion of viscous fluids. These equations can be written as a hyperbolic-parabolic system of conservation laws in the following manner
\begin{equation}
  \dpn{\vectorsym{u}}{t}+\nabla\cdot\vectorsym{f}(\vectorsym{u},\nabla{\vectorsym{u}})=\vectorsym{0}.
\end{equation}
The vector of conserved quantities is defined by $\vectorsym{u}(\vectorsym{x},t)=\left(\rho,\rho\vectorsym{v},\rho{}E\right)$, where $\rho$ is the mass density, $\vectorsym{v}=\left(v_{x},v_{y},v_{z}\right)$ is the velocity vector and $E$ is the total energy. The physical flux is defined by
\begin{equation}
  \vectorsym{f}(\vectorsym{u},\nabla\vectorsym{u})=
    \vectorsym{f}^{A}(\vectorsym{u})-\vectorsym{f}^{D}(\vectorsym{u},\nabla\vectorsym{u}),
\end{equation}
with the advective and diffusive fluxes given, respectively, by
\begin{equation}
  \vectorsym{f}^{A}(\vectorsym{u})=%
    \begin{pmatrix}
      \rho\,\vectorsym{v}\\
      \rho\,\vectorsym{v}\otimes\vectorsym{v}+p\tensorsym{I}\\
      \vectorsym{v}\left(\rho{}E+p\right)
    \end{pmatrix},\quad
  \vectorsym{f}^{D}(\vectorsym{u},\nabla\vectorsym{u})=%
    \begin{pmatrix}
      0\\
      -\boldsymbol{\tau}\\
      \boldsymbol{\tau}\cdot\vectorsym{v}-\vectorsym{q}
    \end{pmatrix}.
\end{equation}
Viscous stresses are described by the stress tensor $\boldsymbol{\tau}$, defined by
\begin{equation}
  \boldsymbol{\tau}=2\mu{}\tensorsym{S}^{D}=\mu\left(\nabla\vectorsym{v}+\left(\nabla\vectorsym{v}\right)^{\intercal}-\frac{2}{3}\left(\nabla\cdot\vectorsym{v}\right)\tensorsym{I}\right),
\end{equation}
where $\tensorsym{S}^{D}$ is the deviatoric component of the strain-rate tensor
\begin{equation}
  \tensorsym{S}=\frac{1}{2}\left(\nabla\vectorsym{v}+\left(\nabla\vectorsym{v}\right)^{\intercal}\right),
\end{equation}
and
\begin{equation}
  \vectorsym{q}=-k\nabla{T},\quad k=\mu\frac{c_{p}}{\mathrm{Pr}}.
\end{equation}
The equation of state is given by
\begin{equation}
  p=\left(\gamma-1\right)\left(\rho{}E-\frac{1}{2}\rho{}v^{2}\right),
\end{equation}
where $p$ is the static pressure, $\gamma$ is the ideal gas index, $\mu$ is the dynamic viscosity, $T$ is the temperature, $k$ is the thermal conductivity and $\mu$ denotes the dynamic viscosity coefficient. The kinematic viscosity coefficient is defined by the formula
\begin{equation}
  \nu = \mu/\rho.
\end{equation}
\subsection{Reynolds-averaged Navier--Stokes equations}\label{sec:governing-equations:rans}
The Navier--Stokes equations govern the motion of fluids in both laminar and turbulent regimes. Because of the large range of spatial and temporal scales present in turbulent flows, directly solving the Navier--Stokes equations is prohibitively expensive. Therefore, the Reynolds--averaged Navier--Stokes (RANS) equations are solved instead to model steady turbulent mean flows. The RANS equations couple the mean flow equations with the one-equation turbulence model of Spallart--Allmaras. These equations can be written also as a hyperbolic-parabolic system of conservation laws with source term in the following way
\begin{equation}
  \dpn{\vectorsym{u}}{t}+\nabla\cdot\vectorsym{f}(\vectorsym{u},\nabla{\vectorsym{u}})=\vectorsym{s}\left(\vectorsym{u},\nabla\vectorsym{u}\right).
\end{equation}
Here, $\vectorsym{u}$ is the vector of time-averaged conservative variables over a given time interval and has the following components: $\vectorsym{u}(\vectorsym{x},t)=\left(\rho,\rho\vectorsym{v},\rho{}E,\rho\tilde{\nu}\right)$, where $\rho$ is the time-averaged mass density, $\vectorsym{v}=\left(v_{x},v_{y},v_{z}\right)$ is the time-averaged velocity vector, $E$ is the time-averaged total energy, and $\rho\tilde{\nu}$ is a new conservative variable relating the time-averaged mass density and the eddy viscosity $\tilde{\nu}$. The advective and diffusive components of the physical flux associated with the quantity $\rho\tilde{\nu}$ are given by
\begin{equation}
  \vectorsym{f}^{A}\left[\rho\tilde{\nu}\right]=\rho\tilde{\nu}\vectorsym{v},\quad
  \vectorsym{f}^{D}\left[\rho\tilde{\nu}\right]=\frac{1}{\sigma}\left(\mu+f_{n1}\rho\tilde{\nu}\right)\nabla{}\tilde{\nu}.
\end{equation}
The turbulent stress tensor $\boldsymbol{\tau}_{t}$ and the turbulent heat fluxes $\vectorsym{q}_{t}$ are added, respectively, to $\boldsymbol{\tau}$ and $\vectorsym{q}$ appearing in the mean flow equations:
\begin{equation}
  \boldsymbol{\tau}_{t}=2\mu_{t}\tensorsym{S}^{D},\quad
  \vectorsym{q}_{t}=-\frac{\mu_{t}}{\mathrm{Pr}_{t}}c_{p}\nabla{}T,
\end{equation}
where $\mathrm{Pr}_{t}=0.9$ is the turbulent Prandtl number and $\mu_{t}$ is the turbulent dynamic viscosity
\begin{equation}
  \mu_{t}=
    \begin{cases}
      \rho\tilde{\nu}f_{v1}\left(\chi\right) & \text{for $\tilde{\nu}\ge{}0$},\\
      0 & \text{for $\tilde{\nu}<0$},\\
    \end{cases}
\end{equation}
with
\begin{equation}
  f_{v1}\left(\chi\right)=\frac{\chi^{3}}{\chi^{3}+c_{v1}^{3}},\quad
  \chi=\frac{\rho\tilde{\nu}}{\mu}.
\end{equation}
The source terms only have a nonzero component in the equation for the turbulent variable $\rho\tilde{\nu}$:
\begin{equation}
  \vectorsym{S}\left[\rho\tilde{\nu}\right]=-\rho\left(P-D\right)-\frac{c_{b2}}{\sigma}\rho\nabla\tilde{\nu}+\frac{1}{\sigma}\left(\nu+f_{n1}\tilde{\nu}\right)\nabla\rho\cdot\nabla\tilde{\nu},
\end{equation}
where the production and destruction terms, $P$ and $D$, are defined by:
\begin{align}
  P&=
    \begin{cases}
      c_{b1}\left(1-f_{t2}\right)\tilde{\omega}\tilde{\nu} & \text{for $\tilde{\nu}\ge{}0$},\\
      c_{b1}\left(1-c_{t3}\right)\omega\tilde{\nu} & \text{for $\tilde{\nu}<0$},\\
    \end{cases}\\
  D&=
    \begin{cases}
      \left(c_{w1}f_{w}-\frac{c_{b1}}{\kappa^{2}}f_{t2}\right)\left(\frac{\tilde{\nu}}{d}\right)^{2}
       & \text{for $\tilde{\nu}\ge{}0$},\\
      -c_{w1}\left(\frac{\tilde{\nu}}{d}\right)^{2} & \text{for $\tilde{\nu}<0$},\\
    \end{cases}
\end{align}
and
\begin{align}
  f_{n1}&=
    \begin{cases}
      1 & \text{for $\tilde{\nu}\ge{}0$},\\
      \frac{c_{n1}+\chi^{3}}{c_{n1}-\chi^{3}} & \text{for $\tilde{\nu}<0$},\\
    \end{cases}\\
  f_{t2}&=c_{t3}\exp{\left(-c_{t4}\chi^{2}\right)},\\
  f_{w}&=g\left(\frac{1+c_{w3}^{6}}{g^{6}+c_{w3}^{6}}\right)^{\frac{1}{6}},
\end{align}
with
\begin{equation}
  g=r+c_{w2}\left(r^{6}-r\right),\quad r=\min\left(r_{\mathrm{lim}},\frac{\tilde{\nu}}{\omega\kappa^{2}d^{2}}\right),
\end{equation}
and $d$ is the distance to the nearest wall and $\omega$ is the magnitude of the vorticity. The modified vorticity magnitude $\tilde{\omega}$ is given by
\begin{equation}
  \tilde{\omega}=
    \begin{cases}
      \omega+\bar{\omega} & \text{for $\bar{\omega}>-c_{v2}\omega$},\\
      \omega+\frac{\omega\left( c_{v2}^{2}+c_{v3}\bar{\omega}\right)}{\left(c_{v3}-2c_{v2}\right)\omega-\bar{\omega}} & \text{for $\bar{\omega}<-c_{v2}\omega$},
    \end{cases}
\end{equation}
where $\bar{\omega}$ and $f_{v2}$ are given by
\begin{equation}
  \bar{\omega}=\frac{\tilde{\nu}}{\kappa^{2}d^{2}}f_{v2},\quad f_{v2}=1-\frac{\chi}{1-f_{v1}}.
\end{equation}
For completeness, we give the values of the constants in the above expressions:
${c_{v1}=\num{7.1}}$,
${\sigma=2/3}$,
${c_{b1}=\num{0.1355}}$,
${c_{b2}=\num{0.622}}$,
${\kappa=\num{0.41}}$,
${c_{w2}=\num{0.3}}$,
${c_{w3}=2}$,
${r_{\mathrm{lim}}=\num{10}}$,
${c_{t3}=1.2}$,
${c_{t4}=0.5}$,
${c_{v2}=0.7}$,
${c_{v3}=0.9}$,
${c_{n1}=16}$.
%
\section{Numerical methods}\label{sec:numerical-methods}
The spatial discretization schemes used in CODA to solve the Navier--Stokes and Reynolds-Averaged Navier--Stokes equations are the finite volume method, the modal discontinuous Galerkin method, and the discontinuous Galerkin spectral element method. These numerical schemes are implemented in order to solve partial differential equations in a multi-physics context. Here we will briefly describe the finite volume and modal discontinuous Galerkin methods for hyperbolic systems of conservation laws. Regarding the time discretization, in CODA there are available explicit, implicit, and implicit-explicit integrators. In this work, we have used an implicit backward-Euler scheme based on preconditioned matrix-free GMRES.\par
As we have already seen, the partial differential equations of interest can be written as a hyperbolic-parabolic system of conservation laws with source terms in differential form
\begin{equation}\label{eqn:conservation-law:differential-form}
  \dpn{\vectorsym{u}}{t}+\nabla\cdot\vectorsym{f}(\vectorsym{u},\nabla{\vectorsym{u}})=\vectorsym{s}\left(\vectorsym{u},\nabla\vectorsym{u}\right),
\end{equation}
where $\vectorsym{u}=\vectorsym{u}(\vectorsym{x},t)$ is the vector of conserved quantities and $\vectorsym{f}=\vectorsym{f}(\vectorsym{u},\nabla\vectorsym{u})$ is the tensor of physical fluxes, which has two terms, the advective term and the diffusive term, namely
\begin{equation}
  \vectorsym{f}(\vectorsym{u},\nabla\vectorsym{u})=
    \vectorsym{f}^{A}(\vectorsym{u})-\vectorsym{f}^{D}(\vectorsym{u},\nabla\vectorsym{u}),
\end{equation}
and $\vectorsym{s}=\vectorsym{s}\left(\vectorsym{u},\nabla\vectorsym{u}\right)$ represents the source terms. The \cref{eqn:conservation-law:differential-form} can also be written in integral form as follows
\begin{equation}\label{eqn:conservation-law:integral-form}
  \dpn{}{t}\int_{\Omega}\vectorsym{u}\diff{\Omega} +
  \int_{\partial\Omega}\vectorsym{f}(\vectorsym{u},\nabla\vectorsym{u})\cdot\vectorsym{n}\diff{\sigma} = \int_{\Omega}\vectorsym{s}\left(\vectorsym{u},\nabla\vectorsym{u}\right)\diff{\Omega}.
\end{equation}
Here, $\Omega$ is the spatial domain and $\vectorsym{n}$ is the normal vector on the surface $\partial\Omega$ enclosing the spatial domain. The numerical methods we discuss in this section are based on \cref{eqn:conservation-law:integral-form}.
\subsection{Finite volume methods}\label{sec:numerical-methods:finite-volume-methods}
In the finite-volume framework~\cite{blazek2015a}, the computational domain is divided into a set of non-overlapping polyhedral control volumes or cells, that is, $\Omega=\bigcup_{i}^{N}\Omega_{k}$, and the integral form of the equations is discretized for each control volume $\Omega_{k}$. Defining the mean value of the function $\vectorsym{u}$ in the cell $\Omega_{k}$ by
\begin{equation}
  \vectorsym{u}_{k}:=\dfrac{1}{\left|\Omega_{k}\right|}\int_{\Omega_{k}}\vectorsym{u}\left(\vectorsym{x},t\right)\diff{\Omega},
\end{equation}
we get the so-called semi-discrete formulation
\begin{equation}
  \dn{\vectorsym{u}_{k}}{t}:=-\dfrac{1}{\left|\Omega_{k}\right|}
    \biggl\{
    \sum_{l}\bigl[\vectorsym{f}^{A}\cdot\hat{\vectorsym{n}}\bigr]^{*}\sigma_{l}-\sum_{l}\bigl[\vectorsym{f}^{D}\cdot\hat{\vectorsym{n}}\bigr]^{*}\sigma_{l}
    - \vectorsym{s}\left|\Omega_{k}\right|
    \biggr\}.
\end{equation}
The surface integral has been approximated by a sum of fluxes through the faces $\partial\Omega_{l}$ of the cell $\Omega_{k}$. The $\sigma_{l}$ stands for the area of the face $\partial\Omega_{l}$, $\hat{\vectorsym{n}}$ is the normal vector of the face $\partial\Omega_{l}$, and $\left|\Omega_{k}\right|$ is the volume of the cell $\Omega_{k}$. The fluxes across the faces of the elements are calculated by numerical fluxes (this is represented in the equation as the operator $\left[\cdot\right]^{*}$). The computation of the numerical advective fluxes requires a reconstruction procedure that takes cell-centered values of conserved quantities, momentum and temperature gradients, and interpolates them to the faces. Typically, Roe schemes with an entropy fix are employed. The gradients are computed using the Green--Gauss theorem. This procedure reconstructs values needed at face integration points to approximate the Green--Gauss integral linearly. Gradient limiters can be applied in the vicinity of strong discontinuities. The finite volume discretization implemented in CODA is second-order accurate. More details on how the element and face gradients are computed can be found in~\citet{blazek2015a}.
\subsection{Discontinuous Galerkin methods}\label{sec:numerical-methods:discontinuous-galerkin-methods}
The discontinuous Galerkin method is based on the weak formulation of the governing equations. The method has the capacity to handle complicated geometries and, due to its locality, is highly parallelizable. Arbitrary high-order discontinuous Galerkin discretizations can be obtained simply by increasing the degree of the polynomial basis functions. In this work, we have used the modal approach of the discontinuous Galerkin scheme, which uses a hierarchical and orthogonal polynomial basis.\par
Let us start by considering the domain $\Omega$ with boundary $\partial\Omega$. We decompose $\Omega$ into a shape-regular partition $\mathcal{T}_{h}=\left\{\Omega_{k}\right\}$ consisting of $N$ non-overlapping elements $\Omega_{k}$ of characteristic size $h$. These elements can be, for instance, tetrahedra, hexahedra, pyramids, and prisms. Let us define the sets $\mathcal{F}_{i}$ and $\mathcal{F}_{b}$ of the interior and boundary faces in $\mathcal{T}_{h}$, such that $\mathcal{F}_{h}=\mathcal{F}_{i}\cup\mathcal{F}_{b}$.\par
Let $\mathcal{V}_{h}^{p}=\left\{\phi_{h}\in L^{2}(\Omega)\colon\ \phi|_{\Omega_{k}}\in\mathcal{P}^{p}(\Omega_{k}),\ \forall{}\Omega_{k}\in\mathcal{T}_{h}\right\}$ be the functional space formed by piecewise polynomials of degree at most $p$ in each element $\Omega_{k}$, and $\left\{\phi_{\Omega_{k}}^{1},\ldots,\phi_{\Omega_{k}}^{N_{p}}\right\}\in\mathcal{P}^{p}(\Omega_{k})$ a hierarchical and orthonormal basis of $\mathcal{V}_{h}^{p}$, of dimension $N_{p}$. As a hierarchical and orthonormal basis, we used one computed with the methodology introduced by~\citet{bassi2012a}, where an initial set of monomial basis functions is defined in each element arbitrarily shaped, and then a modified Gram--Schmidt orthonormalization procedure is applied. In each element of the discretization, the resulting basis yields a diagonal mass matrix that simplifies the resolution of the equations in the variational formulation. Using this basis, the number of degrees of freedom per element is $N_{p}=(p+1)(p+2)(p+3)/6$ in a three-dimensional space. The solution in each element $\Omega_{k}\in\mathcal{T}_{h}$ and for all $t\ge{0}$ is
\begin{equation}
  \vectorsym{u}_{h}(\vectorsym{x},t)=\sum_{l=1}^{N_{p}}\phi_{\Omega_{k}}^{l}(\vectorsym{x})\vectorsym{u}_{\Omega_{k}}^{l}(t),\ \forall\vectorsym{x}\in{}\Omega_{k}.
\end{equation}
Polynomial coefficients $\vectorsym{u}_{\Omega_{k}}^{l}$, for ${1\le{}l\le{}N_{p}}$, are the degrees of freedom of the discrete problem in the element $\Omega_{k}\in\mathcal{T}_{h}$.
The system of conservation laws with source terms given in \cref{eqn:conservation-law:differential-form} can also be written in a weak form as follows
\begin{equation}\label{eqn:conservation-law:weak-form}
  \dpn{}{t}\int_{\Omega}\phi\vectorsym{u}\diff{\Omega}
  -\int_{\Omega}\nabla\phi\cdot\vectorsym{f}(\vectorsym{u},\nabla{\vectorsym{u}})\diff{\Omega}
  +\int_{\partial\Omega}\phi\vectorsym{f}(\vectorsym{u},\nabla{\vectorsym{u}})\cdot\vectorsym{n}\diff{\sigma}
  =\int_{\Omega}\phi\vectorsym{s}(\vectorsym{u},\nabla{\vectorsym{u}})\diff{\Omega},
\end{equation}
where $\phi$ is an arbitrary, sufficiently smooth, test funtion. Because the discontinuous nature of the numerical solution at cell interfaces, the advective and diffusive fluxes require a special treatment. Typically, the advective flux is discretized by using a numerical flux such that it is consistent and conservative, and accounts for wave propagation, for example, the Roe flux. The diffusive flux is discretized using the \textrm{BR2} scheme of~\citet{bassi1997b} in order to obtain a consistent, stable and accurate discretization. This methodology is based on the definition of the gradients of conservative variables as auxiliary variables $\boldsymbol{\Theta}=\nabla\vectorsym{u}$, which verify the following equations:
\begin{align}
  &\boldsymbol{\Theta}-\nabla\vectorsym{u}=\vectorsym{0},\\
  &\dpn{\vectorsym{u}}{t}+\nabla\cdot\vectorsym{f}\left(\vectorsym{u},\boldsymbol{\Theta}\right)=\vectorsym{s}\left(\vectorsym{u},\boldsymbol{\Theta}\right).
\end{align}
For a given cell interface between two elements, we define the average operator as
\begin{equation}
  \average{\vectorsym{u}}=\frac{1}{2}\left(\vectorsym{u}^{+}+\vectorsym{u}^{-}\right),
\end{equation}
and the jump operator as
\begin{equation}
  \jump{\vectorsym{u}}=\vectorsym{u}^{+}\otimes\vectorsym{n}-\vectorsym{u}^{-}\otimes\vectorsym{n},
\end{equation}
where $\vectorsym{u}^{+}$ and $\vectorsym{u}^{-}$ are the values of the solution $\vectorsym{u}$ at the interface between elements $\Omega_{k}^{+}$ and $\Omega_{k}^{-}$. In this way, the discontinuous Galerkin formulation of~\cref{eqn:conservation-law:weak-form} is given by~\cite{arnold2002a,bassi2011a}
\begin{equation}
  \begin{gathered}
    \dpn{}{t}\int_{\Omega_{k}}\phi_{h}\vectorsym{u}_{h}\diff{\Omega}
    -\int_{\Omega_{k}}\nabla_{h}\phi_{h}\cdot\vectorsym{f}\left(\vectorsym{u}_{h},\nabla_{h}{\vectorsym{u}_{h}}
      +\vectorsym{r}\left(\jump{\vectorsym{u}_{h}}\right)\right)\diff{\Omega}\\
    +\int_{\partial\Omega_{k}}\jump{\phi_{h}}\cdot\hat{\vectorsym{f}}\left(\vectorsym{u}_{h}^{\pm},\left(\nabla_{h}{\vectorsym{u}}_{h}+\eta_{F}\vectorsym{r}_{F}\left(\jump{\vectorsym{u}_{h}}\right)\right)^{\pm}\right)\diff{\sigma}
    =\int_{\Omega_{k}}\phi_{h}\vectorsym{s}\left(\vectorsym{u}_{h},\nabla_{h}{\vectorsym{u}_{h}}+\vectorsym{r}\left(\jump{\vectorsym{u}_{h}}\right)\right)\diff{\Omega}.
  \end{gathered}
\end{equation}
For all $F\in\mathcal{F}_{h}$, the local lifting operator $\vectorsym{r}_{F}$ is defined as the solution of the following problem
\begin{equation}
  \int_{\Omega_{k}}\phi_{h}\cdot\vectorsym{r}_{F}\left(\vectorsym{v}\right)\diff{\Omega}
    =-\int_{F}\average{\phi_{h}}\cdot\vectorsym{v}\diff{\sigma},
\end{equation}
for all $\phi_{h}\in\mathcal{V}_{h}^{p}$, and for all $\vectorsym{v}\in\left[L^{2}(F)\right]^{d}$. The global lifting operator $\vectorsym{r}$ is related to $\vectorsym{r}_{F}$ by the equation
\begin{equation}
  \vectorsym{r}\left(\vectorsym{v}\right)=\sum_{F\in\mathcal{F}_{h}}\vectorsym{r}_{F}\left(\vectorsym{v}\right).
\end{equation}
The local lifting operators on the two sides of any face $F$ have support on the two elements sharing the face $F$. Therefore, the global lifting operator for any element $\Omega_{k}\in\mathcal{T}_{h}$ has support on the element $\Omega_{k}$ itself and on its neighbors. As we have already mentioned, the advective and diffusive parts of the numerical flux $\hat{\vectorsym{f}}$ are discretized independently. The numerical advective flux is typically aproximated with the Roe flux. In contrast, the numerical diffusive flux is given by~\cite{arnold2002a,bassi2011a}
\begin{equation}
  \hat{\vectorsym{f}}^{D}\left(\vectorsym{u}_{h}^{\pm},\left(\nabla_{h}{\vectorsym{u}}_{h}+\eta_{F}\vectorsym{r}_{F}\left(\jump{\vectorsym{u}_{h}}\right)\right)^{\pm}\right)=
  \average{\vectorsym{f}^{D}\left(\vectorsym{u}_{h},\nabla_{h}\vectorsym{u}_{h}+\eta_{F}\vectorsym{r}_{F}\left(\jump{\vectorsym{u}_{h}}\right)\right)},
\end{equation}
where the stability parameter $\eta_{F}$ must be greater than the number of faces of the elements~\cite{arnold2002a}. The BR2 diffusive flux discretization is as compact as possible because it keeps the computation strictly bounded to immediate face-neighboring elements, which significantly saves computational memory and enhances execution parallelization.\par
\subsection{Backward Euler method}\label{sec:numerical-methods:backward-euler-method}
Regarding the time discretization, we have employed an implicit scheme. Because this study focuses strictly on steady-state solutions of flow problems modeled by the Navier--Stokes and Reynolds-Averaged Navier--Stokes (RANS) equations, we employed the linearized implicit Backward Euler scheme. For steady-state simulations, high-order temporal accuracy is unnecessary since the time derivative approaches zero at convergence. Instead, the selection criteria prioritize solver efficiency and robustness. The implicit Backward Euler integrator is unconditionally stable and highly robust, features low memory overhead, and parallelizes efficiently. To ensure stability and accelerate convergence to the steady state, the CODA CFD solver utilizes a local pseudo-time-stepping algorithm. During the initial iterations, the solver maintains a small global Courant--Friedrichs--Lewy (CFL) number to handle strong initial transients safely, which is subsequently increased to accelerate convergence. While higher-order time integrators could technically be applied, their additional computational expense offers no benefit for steady-state verification.\par
When the method of lines is applied to \cref{eqn:conservation-law:integral-form}, we get, for each element $\Omega_{k}$, a system of coupled ordinary differential equations in time
\begin{equation}\label{eqn:method-of-lines}
  \matrixsym{M}\dn{\vectorsym{u}_{k}}{t}=-\vectorsym{R}\left(\vectorsym{u}_{k}\right),
\end{equation}
where $\vectorsym{u}_{k}$ is the vector of conservative variables in element $\Omega_{k}$, $\matrixsym{M}$ is the mass matrix, and $\vectorsym{R}$ is the residual, that is, the complete spatial discretization including the source terms.\par
Approximating \cref{eqn:method-of-lines} with the backward Euler method yields
\begin{equation}\label{eqn:backward-euler:base}
  \matrixsym{M}\dfrac{\Delta\vectorsym{u}^{n}_{k}}{\Delta{}t}=-\vectorsym{R}\left(\vectorsym{u}^{n+1}_{k}\right),
\end{equation}
where
\begin{equation}
  \Delta\vectorsym{u}^{n}_{k}=\vectorsym{u}^{n+1}_{k}-\vectorsym{u}^{n}_{k}.
\end{equation}
The implicit formulation leads to a set of non-linear equations for the unknown flow variables at the time $t^{n+1}$. The residual $\vectorsym{R}\left(\vectorsym{u}^{n+1}_{k}\right)$ is linearized according to
\begin{equation}\label{eqn:backward-euler:linearization}
  \vectorsym{R}\left(\vectorsym{u}^{n+1}_{k}\right)=\vectorsym{R}\left(\vectorsym{u}^{n}_{k}\right)+\left(\dpn{\vectorsym{R}}{\vectorsym{u}}\right)^{n}_{k}\Delta\vectorsym{u}^{n}_{k},
\end{equation}
where $\left(\dpn{\vectorsym{R}}{\vectorsym{u}}\right)^{n}_{k}$ is the flux Jacobian matrix. If we substitute the linearization~\cref{eqn:backward-euler:linearization} into~\cref{eqn:backward-euler:base}, we get the implicit scheme
\begin{equation}\label{eqn:backward-euler:scheme}
  \left[\dfrac{\matrixsym{M}}{\Delta{}t} + \left(\dpn{\vectorsym{R}}{\vectorsym{u}}\right)^{n}_{k}\right]\Delta\vectorsym{u}^{n}_{k}=-\vectorsym{R}\left(\vectorsym{u}^{n}_{k}\right).
\end{equation}
\subsection{Immersed boundary methods}\label{sec:numerical-methods:immersed-boundary-methods}
The immersed boundary method was introduced by~\citet{peskin1972a} to analyze flow through the native mitral heart valve, and since then the method has been applied to the numerical simulation of flow problems in complex geometries~\cite{iaccarino2003a}, multiphase flows~\cite{shao2013a} and fluid-structure interaction~\cite{wang2015d}.\par
Immersed boundary methods are numerical techniques used to simulate fluid-structure interaction problems involving complex geometries, providing a flexible framework for modeling fluid flow over complicated immersed bodies on meshes that do not conform to the surface of the body by treating these as immersed boundaries within the computational domain, and therefore, an alternative to numerical methods based on body-fitted meshes. In this way, immersed boundary methods overcome the challenges of generating high-quality meshes around complex geometries, which can be time-consuming and computationally demanding, while accurately resolving complex flows using simple grids. The geometric components are discretized using a separate representation, such as a surface or volume mesh, which is then immersed within the computational mesh. Special immersed boundary conditions are applied to take into account the geometric components in the flow.\par
In industrial applications, like aerodynamics of aircrafts, a high mesh resolution is necessary in order to accurately model the boundary layer. In these kinds of situation, the use of immersed boundary methods to simulate fluid flow over immersed bodies requires a very high resolution of this background mesh to obtain reliable numerical solutions. Using very fine meshes in the computational domain makes the use of immersed boundary methods for industrial applications unpractical. Actually, fine meshes are only necessary in specific locations in the computational domain, like around the immersed geometry or wake regions. Fine meshes far away from the immersed bodies are unnecessary. This aspect of the physics and numerics of the problem motivates the development of mesh refinement algorithms to refine the mesh only in the regions where it is, in fact, essential for accurate computations.\par
As we have already mentioned, there are different immersed boundary techniques currently available, among them, the more relevant are the cut-cell methods, the ghost cell methods, the direct forcing approach, sharp interface methods, and the volume penalization schemes. The latter is of special interest to us and is employed in this work. In the immersed boundary volume penalization methods, the flow equations are solved on the whole computational domain, while a source term is introduced to represent the body as a porous medium with very low permeability. The main idea of this method is to apply a penalization to the integration points which are located inside the immersed geometry using source terms instead of using a boundary condition on the surface of the body, as is the case when using conforming body-fitted meshes. In this manner, a simple Cartesian mesh can be used to solve the flow equations over arbitrary geometries, reducing the time required for the mesh generation.\par
\begin{figure}
  \centering
  \includegraphics[width=0.75\linewidth]{\figurespath/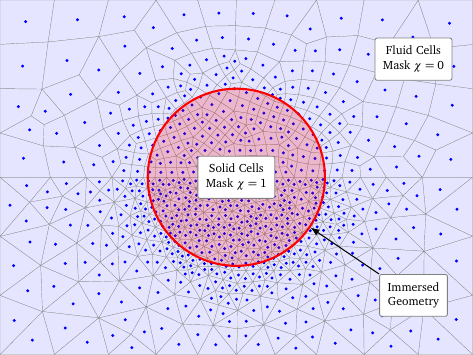}%
  \caption{Schematic diagram of a hybrid background mesh for the immersed boundary method. A mask function is used to discriminate between those points and cells that are inside or outside the immersed body.}%
  \label{fig:numerical-methods:ibm-hybrid-mesh}%
\end{figure}
In the immersed boundary volume penalization method, a mask function is used to discriminate between the points inside or outside the immersed geometry (see \cref{fig:numerical-methods:ibm-hybrid-mesh}). The mask function is computed with the ray casting method~\cite{roth1982a,bonet1991a,ericson2005a,de-berg2008a}. In this approach, a ray is generated starting from a given point and follows any direction; the idea is to test how many times this ray intersects the faces of the immersed geometry. If the point is located outside the body surface, the ray will intersect the faces an even number of times. If the points are located inside the body surface, the ray will intersect the faces an odd number of times. Once the mask function has been calculated in the preprocessing phase of the simulation, the source terms are applied to the flow equations to modify the behavior of the flow field inside the immersed body.\par
Provided the governing equations for a compressible viscous fluid,
\begin{equation}
  \dpn{\vectorsym{u}}{t}+\nabla\cdot\vectorsym{f}(\vectorsym{u},\nabla{\vectorsym{u}})=\vectorsym{s}(\vectorsym{u}),
\end{equation}
the immersed boundary source term $\vectorsym{s}(\vectorsym{u})$ is given by
\begin{equation}
  \vectorsym{s}(\vectorsym{u};\chi,\eta)=%
  \frac{\chi}{\eta}
    \begin{pmatrix}
      0\\
      \rho(\vectorsym{v}-\vectorsym{v}_{\mathrm{s}})\\
      \dfrac{1}{2}\rho{}\left(\vectorsym{v}\cdot\vectorsym{v}-\vectorsym{v}_{\mathrm{s}}\cdot\vectorsym{v}_{\mathrm{s}}\right)
    \end{pmatrix},
\end{equation}
where $\vectorsym{v}_{\mathrm{s}}$ is the velocity of the moving geometry (for a static object $\vectorsym{v}_{\mathrm{s}}=0$) and $\chi$ represents the mask function and distinguishes between the fluid region $\Omega_{\mathrm{f}}$ and the solid region $\Omega_{\mathrm{s}}$:
\begin{equation}
  \chi\left(\vectorsym{x},t\right) =%
  \begin{cases}
    1, & \text{if $\vectorsym{x}\in\Omega_{\mathrm{s}}$},\\
    0, & \text{otherwise},
  \end{cases}
\end{equation}
and $0<\eta\ll{}1$ is the penalization parameter. If turbulent flow is simulated with the Reynolds-averaged Navier--Stokes equations and the Spalart--Allmaras model is used, the eddy viscosity, $\tilde{\nu}$, is set to zero inside the body~\cite{tamaki2017a}:
\begin{equation}
  \vectorsym{s}_{\mathrm{SA}}(\vectorsym{u};\chi,\eta)=-\frac{\chi}{\eta}\tilde{\nu}.
\end{equation}
In our implementation of immersed boundary volume penalization schemes for discontinuous Galerkin methods, the volume forcing is applied directly at the quadrature points. Consequently, if a geometric boundary crosses an element, the forcing function undergoes a sharp jump between the solid and ﬂuid regions near the immersed boundary. This discontinuity can induce spurious oscillations in the resolved flow variables. To mitigate this, the oscillations can be reduced by smoothing the mask function. This smooths the transition between solid and fluid zones, effectively allowing smaller penalty forces for quadrature points near the interface~\cite{angot1999a,abgrall2014a,kolomenskiy2009a,kou2022a}. While this smoothing approach has been shown to be robust, the preservation of high-order accuracy inherently depends on the choice of the penalization parameter, $\eta$. In general, $\eta$ must be sufficiently small to ensure high fidelity to the boundary geometry. However, a very small $\eta$ leads to highly stiff source terms that severely restrict the explicit time step. We would like to emphasize that the primary scope of the present work is the description of a new geometry-adaptive mesh refinement preprocessing tool and its integration with immersed boundary methods. Comprehensive details regarding the algorithmic implementation, verification, and mathematical aspects of the volume penalization schemes can be found in our dedicated prior literature~\citep{kou2022a,nunez2025a}.
%
\section{Mesh refinement}\label{sec:mesh-refinement}
\begin{figure*}
  \centering
  \includegraphics[width=0.49\linewidth]{\figurespath/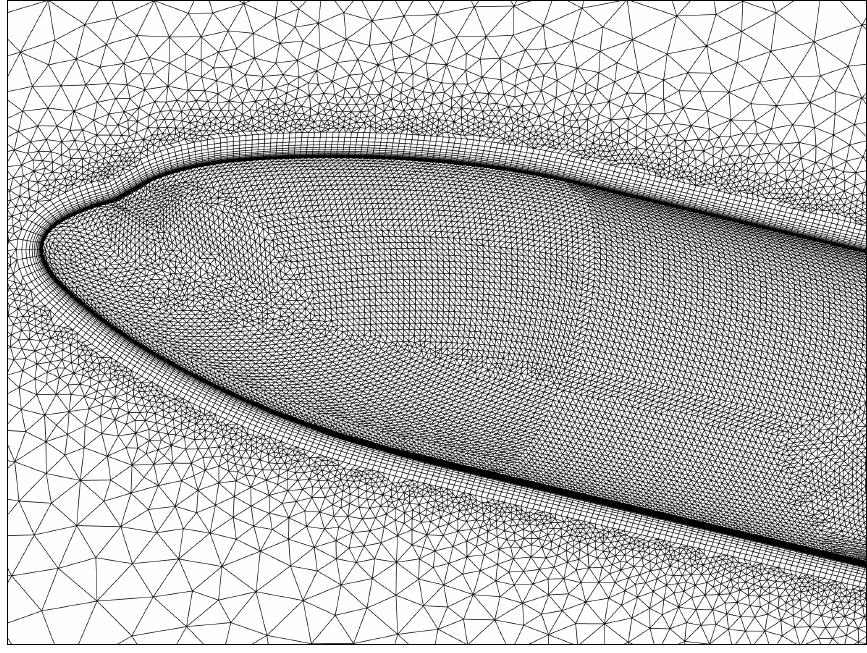}%
  \hfill
  \includegraphics[width=0.49\linewidth]{\figurespath/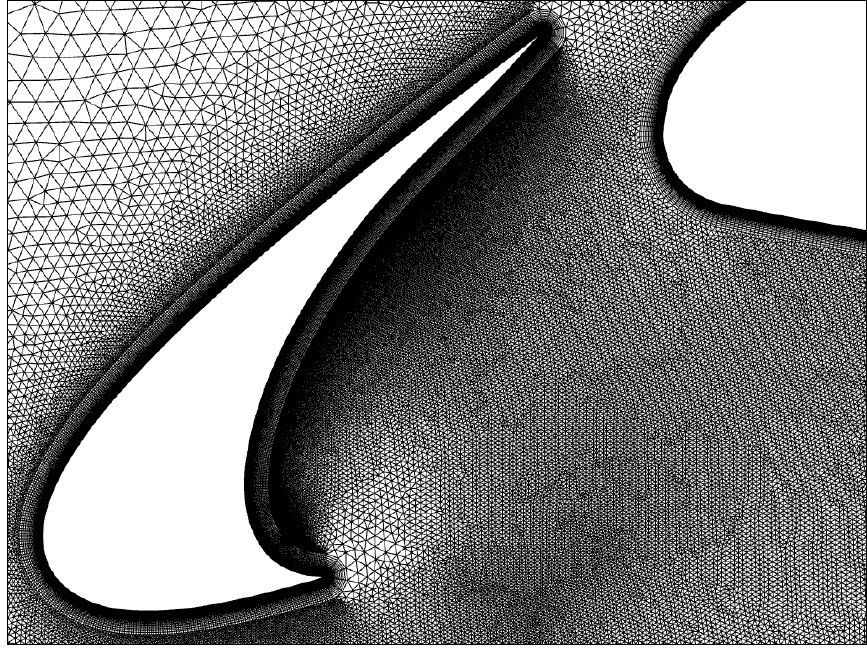}%
  \\[\medskipamount]
  \caption{Hybrid full body-fitted meshes around complex geometries: unstructured grid of a NASA CRM-HL aircraft (left), and unstructured grid of a multi-element airfoil (right).}%
  \label{fig:mesh-refinement:hybrid-body-fitted-mesh}%
\end{figure*}

In this section, we present the mesh refinement algorithms developed and implemented in our preprocessing tool. This software can create an initial Cartesian background mesh made exclusively of hexahedral elements, typically used in immersed-boundary-based simulations. This kind of meshes are very simple to generate, but when they are used in industrial aerodynamic flow problems, a large amount of elements are required if we want to obtain accurate solutions as good as those obtained with full body-fitted meshes. Some industrial simulation problems only require the geometry to be changed in certain specific locations. For instance, if we want to optimize the flap positioning in a multi-element airfoil, it is this new geometry configuration that causes the remeshing of the entire computational domain when using body-fitted meshes. If we want to take advantage of the immersed boundary methods, it would be useful to generate a full body-fitted mesh exclusively for the main airfoil while the flap is ommited. In the region where the flap is supposed to be located, we can place it as an immersed body and refine around its surface (this example is considered in more detail in \cref{sec:numerical-computations}). In order to make feasible this idea, we need that the mesh preprocessing tool be capable of importing an external body-fitted mesh and then refining its elements, which, in general, are of different shapes, like hexahedra, tetrahedra and prisms. Usually hexahedra are used to accurately model the flow on the boundary layer, therefore they are employed on the geometry surface. The rest of the computational domain is filled with prisms and tetrahedra (see \cref{fig:mesh-refinement:hybrid-body-fitted-mesh}).\par
\subsection{Elements splitting convention}
\begin{figure*}
  \centering
  \includegraphics[width=1.0\linewidth]{\figurespath/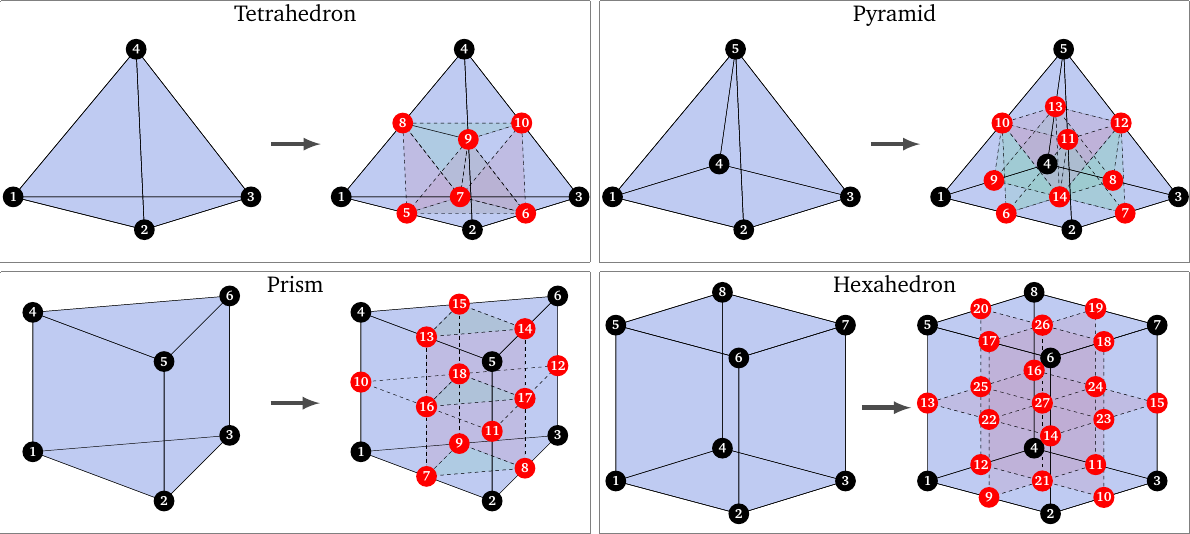}%
  \\[\medskipamount]
  \caption{Splitting of different types of elements implemented in the preprocessing tool. A tetrahedron is split into $8$ tetrahedra. A pyramid is split into $6$ pyramids and $4$ tetrahedra. A prism is split into $8$ prisms. A hexahedron is split into $8$ hexahedra.}%
  \label{fig:mesh-refinement:elements-splitting}%
\end{figure*}

Our mesh refinement preprocessing tool is capable of refining hybrid meshes made of tetrahedra, pyramids, prisms, and hexahedra. In \cref{fig:mesh-refinement:elements-splitting} is depicted the splitting convention adopted in our work, which is typically used in the computational fluid dynamics community. Next, we present the node numbering convention for the children resulting from the element splitting. In the figure hanging nodes can be seen at element faces and edges. These hanging nodes generate nonconforming faces between elements.\par
A tetrahedron is divided into $8$ tetrahedra:
\begingroup
\allowdisplaybreaks
\begin{align*}
  E^{1}_{\mathrm{TETRA4}} &= \{n_{1},n_{5},n_{7},n_{8}\},\\
  E^{2}_{\mathrm{TETRA4}} &= \{n_{5},n_{2},n_{6},n_{9}\},\\
  E^{3}_{\mathrm{TETRA4}} &= \{n_{6},n_{3},n_{7},n_{10}\},\\
  E^{4}_{\mathrm{TETRA4}} &= \{n_{8},n_{9},n_{10},n_{4}\},\\
  E^{5}_{\mathrm{TETRA4}} &= \{n_{5},n_{7},n_{8},n_{9}\},\\
  E^{6}_{\mathrm{TETRA4}} &= \{n_{5},n_{6},n_{7},n_{9}\},\\
  E^{7}_{\mathrm{TETRA4}} &= \{n_{7},n_{8},n_{9},n_{10}\},\\
  E^{8}_{\mathrm{TETRA4}} &= \{n_{7},n_{9},n_{6},n_{10}\}.
\end{align*}
\endgroup
A pyramid is divided into $6$ pyramids and $4$ tetrahedra:
\begingroup
\allowdisplaybreaks
\begin{align*}
  E^{1}_{\mathrm{PYRA5}}   &= \{n_{1},n_{6},n_{14},n_{9},n_{10}\},\\
  E^{2}_{\mathrm{PYRA5}}   &= \{n_{6},n_{2},n_{7},n_{14},n_{11}\},\\
  E^{3}_{\mathrm{PYRA5}}   &= \{n_{9},n_{14},n_{8},n_{4},n_{13}\},\\
  E^{4}_{\mathrm{PYRA5}}   &= \{n_{14},n_{7},n_{3},n_{8},n_{12}\},\\
  E^{5}_{\mathrm{PYRA5}}   &= \{n_{10},n_{11},n_{12},n_{13},n_{5}\},\\
  E^{6}_{\mathrm{PYRA5}}   &= \{n_{10},n_{13},n_{12},n_{11},n_{14}\},\\
  E^{7}_{\mathrm{TETRA4}}  &= \{n_{6},n_{10},n_{11},n_{14}\},\\
  E^{8}_{\mathrm{TETRA4}}  &= \{n_{7},n_{11},n_{12},n_{14}\},\\
  E^{9}_{\mathrm{TETRA4}}  &= \{n_{8},n_{12},n_{13},n_{14}\},\\
  E^{10}_{\mathrm{TETRA4}} &= \{n_{9},n_{13},n_{10},n_{14}\}.
\end{align*}
\endgroup
A prism is divided into $8$ prisms:
\begingroup
\allowdisplaybreaks
\begin{align*}
  E^{1}_{\mathrm{PRISM6}} &= \{n_{1},n_{7},n_{9},n_{10},n_{16},n_{18}\},\\
  E^{2}_{\mathrm{PRISM6}} &= \{n_{7},n_{2},n_{8},n_{16},n_{11},n_{17}\},\\
  E^{3}_{\mathrm{PRISM6}} &= \{n_{9},n_{8},n_{3},n_{18},n_{17},n_{12}\},\\
  E^{4}_{\mathrm{PRISM6}} &= \{n_{7},n_{8},n_{9},n_{16},n_{17},n_{18}\},\\
  E^{5}_{\mathrm{PRISM6}} &= \{n_{10},n_{16},n_{18},n_{4},n_{13},n_{15}\},\\
  E^{6}_{\mathrm{PRISM6}} &= \{n_{16},n_{11},n_{17},n_{13},n_{5},n_{14}\},\\
  E^{7}_{\mathrm{PRISM6}} &= \{n_{18},n_{17},n_{12},n_{15},n_{14},n_{6}\},\\
  E^{8}_{\mathrm{PRISM6}} &= \{n_{16},n_{17},n_{18},n_{13},n_{14},n_{15}\}.
\end{align*}
\endgroup
A hexahedron is divided into $8$ hexahedra:
\begingroup
\allowdisplaybreaks
\begin{align*}
  E^{1}_{\mathrm{HEXA8}} &= \{n_{1},n_{9},n_{21},n_{12},n_{13},n_{22},n_{27},n_{25}\},\\
  E^{2}_{\mathrm{HEXA8}} &= \{n_{9},n_{2},n_{10},n_{21},n_{22},n_{14},n_{23},n_{27}\},\\
  E^{3}_{\mathrm{HEXA8}} &= \{n_{12},n_{21},n_{11},n_{4},n_{25},n_{27},n_{24},n_{16}\},\\
  E^{4}_{\mathrm{HEXA8}} &= \{n_{21},n_{10},n_{3},n_{11},n_{27},n_{23},n_{15},n_{24}\},\\
  E^{5}_{\mathrm{HEXA8}} &= \{n_{13},n_{22},n_{27},n_{25},n_{5},n_{17},n_{26},n_{20}\},\\
  E^{6}_{\mathrm{HEXA8}} &= \{n_{22},n_{14},n_{23},n_{27},n_{17},n_{6},n_{18},n_{26}\},\\
  E^{7}_{\mathrm{HEXA8}} &= \{n_{25},n_{27},n_{24},n_{16},n_{20},n_{26},n_{19},n_{8}\},\\
  E^{8}_{\mathrm{HEXA8}} &= \{n_{27},n_{23},n_{15},n_{24},n_{26},n_{18},n_{7},n_{19}\}.
\end{align*}
\endgroup
As a consequence of this splitting, provided that two initially conforming neighboring elements have the same level of refinement, if one of them is further refined, then the sharing face of this element, which can be a triangular or a quadrilateral face (depending on the kind of elements), will eventually present hanging nodes with a refinement level ratio $\text{2:1}$.\par
\begin{figure*}
  \centering
  \includegraphics[width=1.0\linewidth]{\figurespath/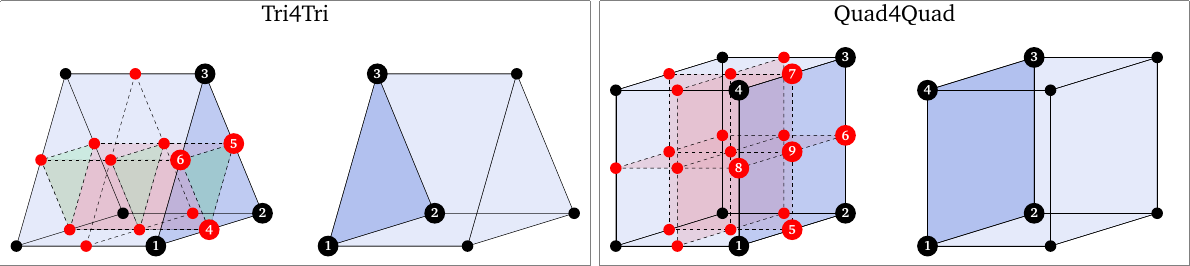}%
  \\[\medskipamount]
  \caption{Hanging nodes numbering convention used by the preprocessing tool for the \textsc{Tri4Tri} (left) and a \textsc{Quad4Quad} (right) interfaces. These interfaces arise between an element with refinement level $l$ and a neighboring element that has been refined up to level $l+1$.}%
  \label{fig:mesh-refinement:nonconforming-interfaces}%
\end{figure*}

The \textsc{Tri4Tri} and \textsc{Quad4Quad} interfaces establish the relationship of the master face nodes with the slave face nodes. We can easily set up arrays of neighboring elements and faces. These arrays connect an element with a refinement level $l$ with its direct neighbors that have been refined to the level $l+1$. We use a convention regarding the nodes numbering to describe the \textsc{Tri4Tri} and \textsc{Quad4Quad} interfaces. For the \textsc{Tri4Tri} interface, if a triangular face with nodes numbers $n_{1}$, $n_{2}$, and $n_{3}$ has been split into four triangular faces and the new hanging nodes created after the split have nodes numbers $n_{4}$, $n_{5}$, and $n_{6}$ (their location is depicted in \cref{fig:mesh-refinement:nonconforming-interfaces}, left), then the numbering of the children faces is as follows: first child face $T_{1}=\{n_{1},n_{4},n_{6}\}$, second child face $T_{2}=\{n_{4},n_{2},n_{5}\}$, third child face $T_{3}=\{n_{6},n_{5},n_{3}\}$, fourth child face $T_{4}=\{n_{4},n_{5},n_{6}\}$. In a similar fashion, for the \textsc{Quad4Quad} interface, if a quadrilateral face with nodes numbers $n_{1}$, $n_{2}$, $n_{3}$, and $n_{4}$ has been split into four quadrilateral faces, and the new hanging nodes created after the split have nodes numbers $n_{5}$, $n_{6}$, $n_{7}$, $n_{8}$, and $n_{9}$ (their location is depicted in \cref{fig:mesh-refinement:nonconforming-interfaces}, right), then the numbering of the children faces is as follows: first child face $Q_{1}=\{n_{1},n_{5},n_{9},n_{8}\}$, second child face $Q_{2}=\{n_{5},n_{2},n_{6},n_{9}\}$, third child face $Q_{3}=\{n_{8},n_{9},n_{7},n_{4}\}$, fourth child face $Q_{4}=\{n_{9},n_{6},n_{3},n_{7}\}$. An efficient algorithm to construct the \textsc{Tri4Tri} and \textsc{Quad4Quad} interfaces plays a key role in the preprocessing tool and is discussed later in this work.\par
For the \textsc{Tri4Tri} interface, the hanging nodes coordinates are computed in the following manner:
\begingroup
\allowdisplaybreaks
  \begin{align}
    n_{4} &= \frac{1}{2}\left(n_{1}+n_{2}\right),\\
    n_{5} &= \frac{1}{2}\left(n_{2}+n_{3}\right),\\
    n_{6} &= \frac{1}{2}\left(n_{1}+n_{3}\right),
  \end{align}
\endgroup
and for the \textsc{Quad4Quad} interface, the hanging nodes coordinates are calculated as follows:
\begingroup
\allowdisplaybreaks
  \begin{align}
    n_{5} &= \frac{1}{2}\left(n_{1}+n_{2}\right),\\
    n_{6} &= \frac{1}{2}\left(n_{2}+n_{3}\right),\\
    n_{7} &= \frac{1}{2}\left(n_{3}+n_{4}\right),\\
    n_{8} &= \frac{1}{2}\left(n_{1}+n_{4}\right),\\
    n_{9} &= \frac{1}{4}\left(n_{1}+n_{2}+n_{3}+n_{4}\right).
  \end{align}
\endgroup
\subsection{Algorithms and data structures}
\begin{figure*}
  \centering
  \includegraphics[width=1.0\linewidth]{\figurespath/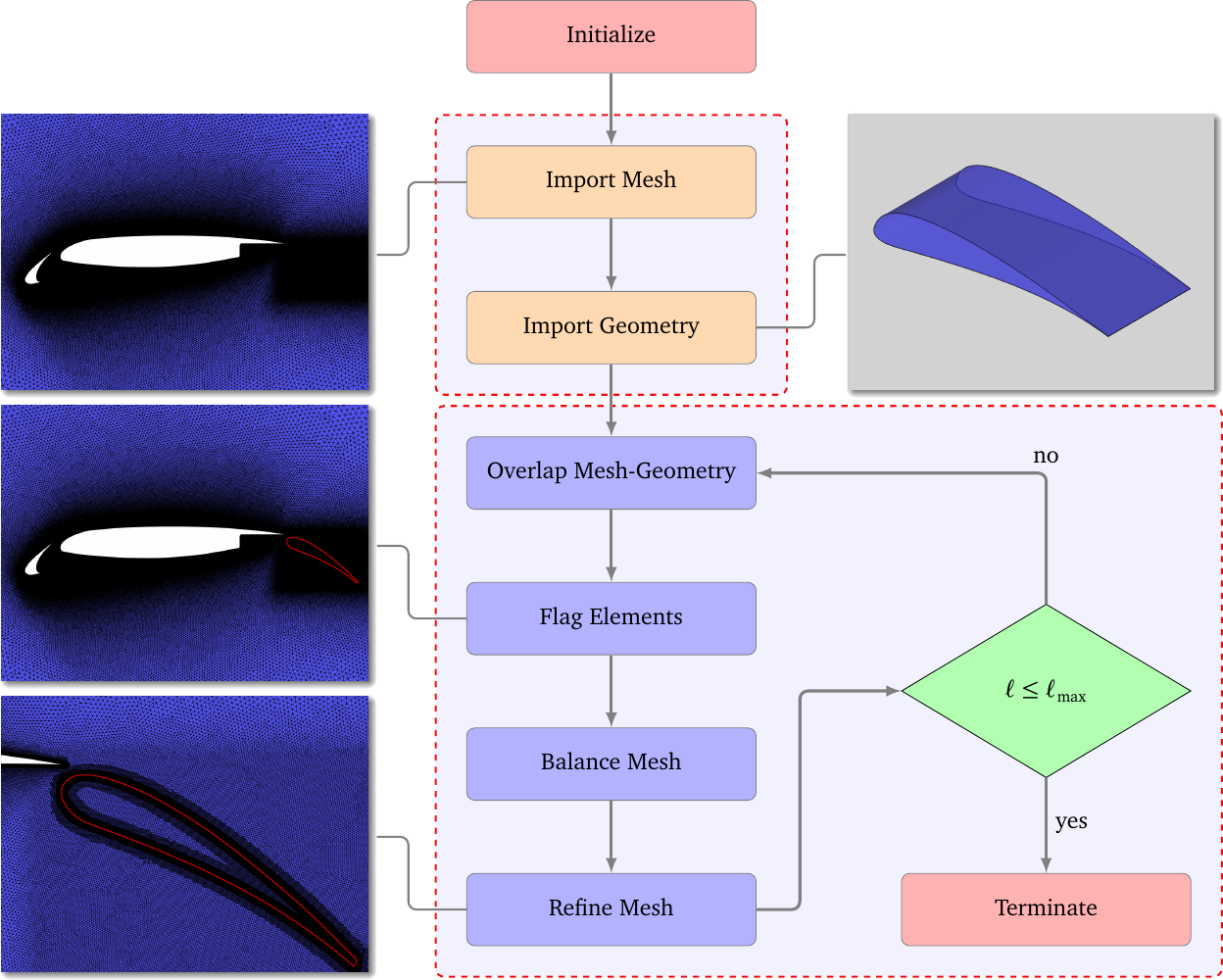}%
  \\[\medskipamount]
  \caption{Flow diagram sketching the necessary steps to refine a hybrid mesh around an immersed geometry.}%
  \label{fig:mesh-refinement:flowchart}%
\end{figure*}

The general idea of the mesh refinement methodology used in this work is sketched in \cref{fig:mesh-refinement:flowchart}. Next we will describe with more detail the algorithms associated with the main refinement loop (\cref{alg:MeshPreprocessing}) and those algorithms used by this, like the distribution of the immersed geometry facets in the mesh (\cref{alg:DistributeGeometryInMesh}), the creation of the arrays of nonconforming sides and elements (\cref{alg:CreateNonConformingSidesList}), the flagging of the elements that overlap with the immersed geometry (\cref{alg:FlagElementsOverlappingGeometry}), the flagging of the elements inside of a box region where we want a finer mesh and where the geometry will be finally immersed (\cref{alg:FlagElementsInsideBox}), the flagging of the elements for balancing such that the refinement level ratio $\text{2:1}$ between neighboring elements is preserved (\cref{alg:FlagElementsForBalancing}), and finally, the refinement of the flagged elements (\cref{alg:RefineElements}).\par
The algorithms implemented in the preprocessing tool work mainly on the following data structures: arrays, linked lists, hash tables, and $k$-d trees. An array of logical values is used to store the flag of elements to be refined (see \textsc{ElementsToRefine} data structure in the algorithms). An array of floats is used to store the vertices coordinates of the facets composing the immersed geometry (see \textsc{GeometryFacets} data structure in the algorithms). The most important variable is the linked list that stores the elements of the mesh (see \textsc{ElementsList} data structure in the algorithms). Each node of this linked list stores, for the corresponding element, its element identification number, its element type, its number of nodes, its refinement flag, its refinement level, the number of facets that overlap with it, a pointer for the faces of the element, a pointer array of the nodes of the element, and pointers to the previous and next element in the linked list. This data structure includes methods for creating, adding, and removing elements, nodes, and faces in the list itself. In this way, full flexibility in the mesh generation is ensured. Hash tables are used mainly to construct mappings between nodes and faces, which allow to find a face if the nodes that constitute this face are provided. A $k$-d tree is used to find the nearest neighbors of the geometry facets to the mesh elements, so that the overlap of the geometry facets with the mesh elements can be efficiently computed.\par
In \cref{alg:MeshPreprocessing} are summarized the main steps to refine a hybrid mesh around an immersed geometry (see \cref{fig:mesh-refinement:flowchart}). The mesh refinement preprocessing tool starts by calling the initialization procedure, where all preprocessing parameters are read from an input file, and then important mesh variables are initialized or allocated. In the second step, the mesh import procedure is called, where an external mesh that has been previously created by, for instance, the GMSH mesh generator, is imported and stored in the \textsc{ElementsList} data structure. In the third step, the geometry import procedure is called, where the surface of the body to be immersed is loaded from a file in STL format and then stored in the \textsc{GeometryFacets} data structure. Next, the \textsc{CreateNonConformingSidesList} procedure (see \cref{alg:CreateNonConformingSidesList}) is called, where the \textsc{NCSList} data structure is created from the \textsc{ElementsList} data structure. It is important to mention that if the mesh is conforming, which means that there are no hanging nodes in this mesh, then the \textsc{Tri4Tri} and \textsc{Quad4Quad} arrays are not created. As mentioned earlier, these arrays allow us to find in a practical manner the neighboring elements with the refinement level $l+1$ of a given master element with the refinement level $l$. This case occurs only if hanging nodes take place. In the next step, a refined subregion is created by calling the \textsc{FlagElementsInsideBox} procedure (see \cref{alg:FlagElementsInsideBox}), where the immersed geometry will be placed. The idea is to have a coarse mesh far away from the zones where the immersed body surfaces are placed and to have a finer mesh close to these surfaces and even finer very close to them. At this point, the immersed geometry has not been inserted into the computational mesh. The region bounding box is defined in the input parameter file, and all elements in the computational domain (provided by the input body-fitted mesh) that lie inside this region are marked as elements to be refined and also are marked as those elements that require a balancing such that the refinement level ratio $\text{2:1}$ between neighboring elements is preserved. This last task is performed by the \textsc{FlagElementsForBalancing} procedure (see \cref{alg:FlagElementsForBalancing}). Once all of these elements have been marked, the \textsc{RefineElements} procedure (see \cref{alg:RefineElements}) is called to refine these elements. This part of the mesh is refined to the level $b_{\mathrm{max}}$. Having refined this box region, refinement of the elements that overlap with the immersed geometry is carried out. In this way, the imported geometry is distributed on the mesh by calling the \textsc{DistributeGeometryInMesh} procedure (see \cref{alg:DistributeGeometryInMesh}), and all elements that intersect the triangles of the body surface are flagged with the \textsc{FlagElementsOverlappingGeometry} procedure (see \cref{alg:FlagElementsOverlappingGeometry}). To preserve the refinement level ratio $\text{2:1}$ between neighboring elements, the flagging of elements is performed to maintain balance, and finally, all marked elements are refined. These steps are repeated until the maximum level of refinement $l_{\mathrm{max}}$ is reached. The resulting mesh is exported in \textrm{HDF5} format, compatible with the CFD solver.\par
\begin{algorithm}{alg:MeshPreprocessing}{MeshPreprocessing}{}
  \begin{algorithmic}[1]
    \Procedure{MeshPreprocessing}{}
      \State {\CallProcedure{Initialization}{}}
      \State {\CallProcedure{ImportBackgroundMesh}{}}
      \State {\CallProcedure{ImportImmersedGeometry}{}}
      \State {\CallProcedure{CreateNonConformingSidesList}{}}

      \For{b}{1}{b_{\mathrm{max}}}
        \State {\CallProcedure{FlagElementsInsideBox}{}}
        \State {\CallProcedure{FlagElementsForBalancing}{}}
        \State {\CallProcedure{RefineElements}{}}
        \State {\CallProcedure{CreateNonConformingSidesList}{}}
      \EndFor

      \For{l}{b_{\mathrm{max}}+1}{l_{\mathrm{max}}}
        \State {\CallProcedure{DistributeGeometryInMesh}{}}
        \State {\CallProcedure{FlagElementsOverlappingGeometry}{}}
        \State {\CallProcedure{FlagElementsForBalancing}{}}
        \State {\CallProcedure{RefineElements}{}}
        \State {\CallProcedure{CreateNonConformingSidesList}{}}
      \EndFor

      \State \CallProcedure{ExportMesh}{}
    \EndProcedure
  \end{algorithmic}
\end{algorithm}
In \cref{alg:DistributeGeometryInMesh} are sketched the steps to distribute the imported geometry (stored in the \textsc{GeometryFacets} data structure) in the imported unstructured hybrid mesh (stored in the \textsc{ElementsList} data structure). To improve the efficiency of the mesh/geometry overlap computation, the geometry facets barycenters are stored in a $k$-d tree data structure (see \textsc{FacetsBary} variable in the algorithms). After creating the $k$-d tree, the intersection of the mesh elements with the geometry facets is tested. For each element, the neighboring facets in the vicinity of the element itself are sought (this vicinity is defined as a sphere of radius ten times the element size; it is advisable that the facets of the geometry have a size of the same order of magnitude as the elements of the refined box region). If an element/facet overlap takes place, the number of facets that intersect the element is stored in the corresponding node of the \textsc{ElementsList} data structure.\par
\begin{algorithm}{alg:DistributeGeometryInMesh}{DistributeGeometryInMesh}{}
  \begin{algorithmic}[1]
    \Procedure{DistributeGeometryInMesh}{}
      \State {\InputVariable{ElementsList}{linked list}}
      \State {\InputVariable{GeometryFacets}{array}}

      \State {\Variable{FacetsBary} $\gets$ \CallFunction{CreateKDTree}{\Variable{GeometryFacets}}}
      \State {\Variable{Elem}~$\to$~\Variable{ElementsList.head}}
      \While {\Variable{Elem}~\algorithmicnot~\algorithmicnil}
        \State {\Variable{BBox} $\gets$ \Variable{Elem.BoundingBox}}
        \State {\Variable{Facet} $\gets$ \CallFunction{NearestNeighbors}{\Variable{BBox},\Variable{FacetsBary}}}

        \If {\CallFunction{TriangleBoxOverlap}{\Variable{BBox},\Variable{Facet}}}
          \State {$\Variable{Elem.nFacets} \gets \Variable{Elem.nFacets}+1$}
        \EndIf

        \State {\Variable{Elem} $\to$ \Variable{Elem.next}}
      \EndWhile

      \State {\ReturnVariable{ElementsList}{linked list}}
    \EndProcedure
  \end{algorithmic}
\end{algorithm}
In \cref{alg:CreateNonConformingSidesList}, the procedure to create the \textsc{Tri4Tri} and \textsc{Quad4Quad} interfaces is explained. The idea is based on the creation of a list with all nonconforming faces of the mesh, that is, the interior faces of the mesh that are not shared between two elements because of the hanging nodes. A face that is shared between two elements shares the same nodes. If hanging nodes are present in the mesh, then two elements can be neighbors while they do not share the same faces (see \cref{fig:mesh-refinement:nonconforming-interfaces}). The goal is to find, for a given master face belonging to an element with refinement level $l$, the neighboring faces belonging to elements with refinement level $l+1$, and that have in common the nodes of the master face. This idea is carried out through the use of hash tables and $k$-d trees data structures. The first step is the creation of nonconforming faces. All interior faces of the mesh belonging to a unique element are stored in the \textsc{NCSList}. This list is used to create a hash table for retrieving a face provided its nodes (see \textsc{NodesToFacesMap} data structure). At the same time, a $k$-d tree of the node coordinates of each nonconforming face is created (see \textsc{NCSNodes}). With these three data structures, the \textsc{Tri4Tri} and \textsc{Quad4Quad} interfaces can be generated. Using the \textsc{NCSList} data structure, the node numbers of a face can be obtained and then the target nodes (these are the hanging nodes) are constructed from these nodes ($n_{4},\ldots,n_{6}$ for a triangular face and $n_{5},\ldots,n_{9}$ for a quadrilateral face); next, the $k$-d tree data structure (\textsc{NCSNodes}) is used to check if the target nodes exist and if they are in the hash table, so the slave faces can be built from them and later store them in the \textsc{Tri4Tri} and \textsc{Quad4Quad} arrays. Once these arrays are constructed, the arrays with the master/slave cell and face numbers relations are created, which will be used in \cref{alg:FlagElementsForBalancing} to mark elements for balancing.\par
\begin{algorithm}{alg:CreateNonConformingSidesList}{CreateNonConformingSidesList}{}
  \begin{algorithmic}[1]
    \Procedure{CreateNonConformingSidesList}{}
      \State {\InputVariable{ElementsList}{linked list}}

      \State {\Variable{NCSList} $\gets$ \CallFunction{CreateNCSides}{\Variable{ElementsList}}}
      \State {\Variable{NodesToFacesMap} $\gets$ \CallFunction{CreateHashTable}{\Variable{NCSList}}}
      \State {\Variable{NCSNodes} $\gets$ \CallFunction{CreateKDTree}{\Variable{NCSList}}}

      \State {\Variable{Side}~$\to$~\Variable{NCSList.head}}
      \While {\Variable{Side}~\algorithmicnot~\algorithmicnil}
        \State {\Variable{NodeIDs} $\gets$ \CallFunction{GetNodeIDs}{\Variable{Side}}}
        \State {\Variable{Target} $\gets$ \CallFunction{CreateTargetNodes}{\Variable{NodeIDs}}}
        \State {\Variable{Found} $\gets$ \CallFunction{NearestNeighbors}{\Variable{NCSNodes},\Variable{Target}}}
        \If {$\Variable{Found}=\Variable{Target}$}
          \State {\Variable{Master} $\gets$ \CallFunction{CreateMasterFaces}{\Variable{NodeIDs}}}
          \State {\Variable{Slaves} $\gets$ \CallFunction{CreateSlaveFaces}{\Variable{NodeIDs},\Variable{Target}}}
          \State {\Variable{MasterSlavesList} $\gets$ \CallFunction{Append}{\Variable{Master},\Variable{Slaves}}}
        \EndIf
        \State {\Variable{Side} $\to$ \Variable{Side.next}}
      \EndWhile
      \State {\Variable{Tri4Tri} $\gets$ \CallFunction{CreateArray}{\Variable{MasterSlavesList}}}
      \State {\Variable{Quad4Quad} $\gets$ \CallFunction{CreateArray}{\Variable{MasterSlavesList}}}

      \State {\ReturnVariable{Tri4Tri}{array}}
      \State {\ReturnVariable{Quad4Quad}{array}}
    \EndProcedure
  \end{algorithmic}
\end{algorithm}
In \cref{alg:FlagElementsOverlappingGeometry} is created the \textsc{ElementsToRefine} array. For each element of the mesh, it is checked if the number of facets intersecting it is greater than zero. If this is the case, the true or false value of this check is stored in the \textsc{ElementsToRefine} array. Recall that in \cref{alg:DistributeGeometryInMesh} this number of facets that overlap with a given element was calculated.
\begin{algorithm}{alg:FlagElementsOverlappingGeometry}{FlagElementsOverlappingGeometry}{}
  \begin{algorithmic}[1]
    \Procedure{FlagElementsOverlappingGeometry}{}
      \State {\InputVariable{ElementsList}{linked list}}

      \State {\Variable{Elem}~$\to$~\Variable{ElementsList.head}}
      \While {\Variable{Elem}~\algorithmicnot~\algorithmicnil}
        \If {$\Variable{Elem.nFacets}>1$}
          \State {\Variable{ElementsToRefine}(\Variable{Elem.ElemID}) $\gets$ \Variable{True}}
        \EndIf
        \State {\Variable{Elem} $\to$ \Variable{Elem.next}}
      \EndWhile

      \State {\ReturnVariable{ElementsToRefine}{array}}
    \EndProcedure
  \end{algorithmic}
\end{algorithm}
In \cref{alg:FlagElementsInsideBox}, as mentioned before, the idea is to mark, for each element of the mesh, if the element is within a subregion previously defined in the parameter file. If this is the case, the true or false value of this test is stored in the  \textsc{ElementsToRefine} array.
\begin{algorithm}{alg:FlagElementsInsideBox}{FlagElementsInsideBox}{}
  \begin{algorithmic}[1]
    \Procedure{FlagElementsInsideBox}{}
      \State {\InputVariable{ElementsList}{linked list}}

      \State {\Variable{BoxRegion} $\gets$ \CallFunction{CreateBoxRegion}{\Variable{ElementsList}}}

      \State {\Variable{Elem}~$\to$~\Variable{ElementsList.head}}
      \While {\Variable{Elem}~\algorithmicnot~\algorithmicnil}
        \If {\CallFunction{IsInsideBox}{\Variable{Elem},\Variable{BoxRegion}}}
          \State {\Variable{ElementsToRefine}(\Variable{Elem.ElemID}) $\gets$ \Variable{True}}
        \EndIf
        \State {\Variable{Elem} $\to$ \Variable{Elem.next}}
      \EndWhile

      \State {\ReturnVariable{ElementsToRefine}{array}}
    \EndProcedure
  \end{algorithmic}
\end{algorithm}
In \cref{alg:FlagElementsForBalancing} are outlined the steps to flag elements in such a way that the refinement level ratio $\text{2:1}$ between neighboring elements is preserved. In order to do this task, it is necessary to use the \textsc{Tri4Tri} and \textsc{Quad4Quad} hanging node interfaces. These interfaces enable us to know the connectivity of neighboring cells with the refinement level $l+1$ of a cell with the refinement level $l$ (see \cref{fig:mesh-refinement:nonconforming-interfaces}). With this information, for each \textsc{Tri4Tri} and \textsc{Quad4Quad} interface, we check that the master element with a refinement level $l-1$ has neighboring refined elements with a refinement level $l$ and they are flagged for refinement. In this way, since already refined elements that have been marked for refinement will become refined elements with refinement level $l+1$, we need to flag the master element, whose refinement level is $l-1$, as an element to be refined, so that the refinement level ratio $\text{2:1}$ between neighboring elements is preserved. If the master element is marked for refinement, then this flag is stored in the \textsc{ElementsToRefine} array.
\begin{algorithm}{alg:FlagElementsForBalancing}{FlagElementsForBalancing}{}
  \begin{algorithmic}[1]
    \Procedure{FlagElementsForBalancing}{}
      \State {\InputVariable{Tri4Tri}{array}}
      \State {\InputVariable{Quad4Quad}{array}}
      \State {\InputVariable{ElementsToRefine}{array}}

      \If {$\Variable{nTri4Tri}>0$}
        \For{l}{L-1}{1}
          \For{\Variable{i}}{1}{\Variable{nTri4Tri}}
            \State {\Variable{Master} $\gets$ \Call{GetMaster}{\Variable{$\Variable{Tri4Tri}(i)$}}}
            \State {\Variable{Slaves} $\gets$ \Call{GetSlaves}{\Variable{$\Variable{Tri4Tri}(i)$}}}
            \State {\Variable{MasterLevel} $\gets$ \Variable{ElementsToLevel}(\Variable{Master})}
            \State {\Variable{Flag1} $\gets$ $\left(\Variable{MasterLevel}=l-1\right)$}
            \State {\Variable{Flag2} $\gets$ $\left(\Call{Any}{\Variable{ElementsToRefine}(\Variable{Slaves}}\right)$}
            \If {\Variable{Flag1}~\algorithmicand~\Variable{Flag2}}
              \State {\Variable{ElementsToRefine}(Master) $\gets$ \Variable{True}}
            \EndIf
          \EndFor
        \EndFor
      \EndIf
      \If {$\Variable{nQuad4Quad}>0$}
        \For{l}{L-1}{1}
          \For{\Variable{i}}{1}{\Variable{nQuad4Quad}}
            \State {\Variable{Master} $\gets$ \Call{GetMaster}{\Variable{$\Variable{Quad4Quad}(i)$}}}
            \State {\Variable{Slaves} $\gets$ \Call{GetSlaves}{\Variable{$\Variable{Quad4Quad}(i)$}}}
            \State {\Variable{MasterLevel} $\gets$ \Variable{ElementsToLevel}(\Variable{Master})}
            \State {\Variable{Flag1} $\gets$ $\left(\Variable{MasterLevel}=l-1\right)$}
            \State {\Variable{Flag2} $\gets$ $\left(\Call{Any}{\Variable{ElementsToRefine}(\Variable{Slaves}}\right)$}
            \If {\Variable{Flag1}~\algorithmicand~\Variable{Flag2}}
              \State {\Variable{ElementsToRefine}(Master) $\gets$ \Variable{True}}
            \EndIf
          \EndFor
        \EndFor
      \EndIf

      \State {\ReturnVariable{ElementsToRefine}{array}}
    \EndProcedure
  \end{algorithmic}
\end{algorithm}
In \cref{alg:RefineElements}, the refinement of the flagged elements is described. The basic idea is to loop over all elements of the mesh (stored in the \textsc{ElementsList} data structure) and check if the element is marked for refinement (stored in the \textsc{ElementsToRefine} data structure). The flagged element is then refined by calling a splitting procedure corresponding to the element type.
\begin{algorithm}{alg:RefineElements}{RefineElements}{}
  \begin{algorithmic}[1]
    \Procedure{RefineElements}{}
      \State {\InputVariable{ElementsList}{linked list}}
      \State {\InputVariable{ElementsToRefine}{array}}

      \State {\Variable{Elem}~$\to$~\Variable{ElementsList.head}}
      \While {\Variable{Elem}~\algorithmicnot~\algorithmicnil}
        \If {\Variable{ElementsToRefine}(\Variable{Elem.ElemID})}
          \Switch {\Variable{Elem.ElemType}}
            \Case {\Variable{TETRA4}}
              \State {\CallProcedure{SplitElement\_TETRA4}{\Variable{Elem}}}
            \EndCase
            \Case {\Variable{HEXA8}}
              \State {\CallProcedure{SplitElement\_HEXA8}{\Variable{Elem}}}
            \EndCase
            \Case {\Variable{PRISM6}}
              \State {\CallProcedure{SplitElement\_PRISM6}{\Variable{Elem}}}
            \EndCase
            \Case {\Variable{PYRA5}}
              \State {\CallProcedure{SplitElement\_PYRA5}{\Variable{Elem}}}
            \EndCase
          \EndSwitch
        \EndIf
        \State {\Variable{Elem} $\to$ \Variable{Elem.next}}
      \EndWhile

      \State {\ReturnVariable{ElementsList}{linked list}}
    \EndProcedure
  \end{algorithmic}
\end{algorithm}
%
\section{Numerical computations}\label{sec:numerical-computations}
In this section, we discuss some test problems in order to illustrate the capabilities of the immersed boundary methods on hexahedral and hybrid unstructured meshes refined by our mesh preprocessing tool. These problems are solved by the CODA CFD solver using the Navier--Stokes and Reynolds-Averaged Navier--Stokes equations. The experimental results of most of these problems are reported in the literature, and thus will serve as validation of our methodology. We have employed the second-order finite volume method and the second- and third-order discontinuous Galerkin method, both schemes coupled with the immersed boundary volume penalization. The diffusive terms in the discontinuous Galerkin method are discretized with the BR2 scheme. A linearized implicit Euler scheme is employed for the time discretization. All simulations were executed using $\num{192}$ cores on the Magerit Supercomputer at the Supercomputing and Visualization Center of Madrid (CeSViMa).\par
\begin{figure*}
  \centering
  \includegraphics[width=0.325\linewidth]{\figurespath/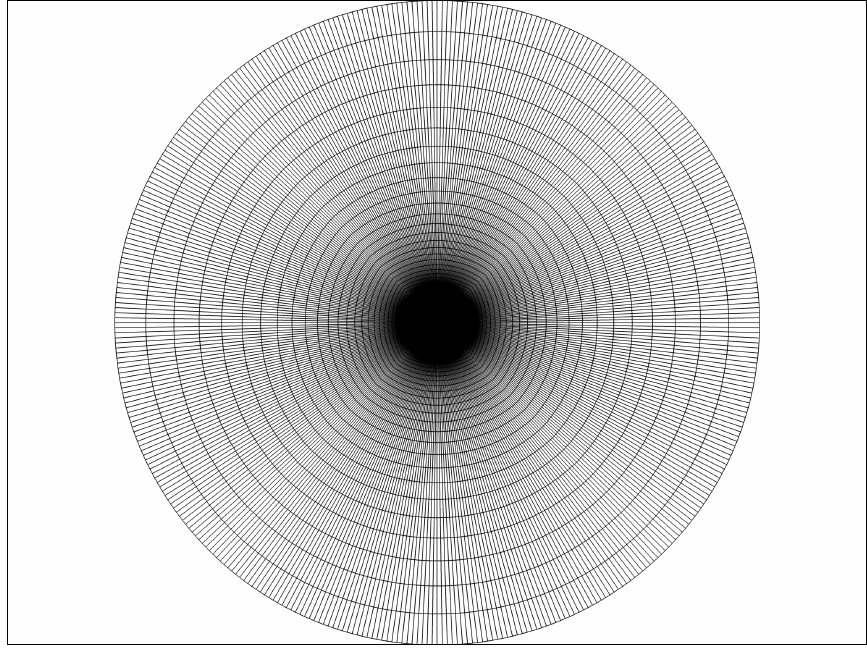}%
  \hfill%
  \includegraphics[width=0.325\linewidth]{\figurespath/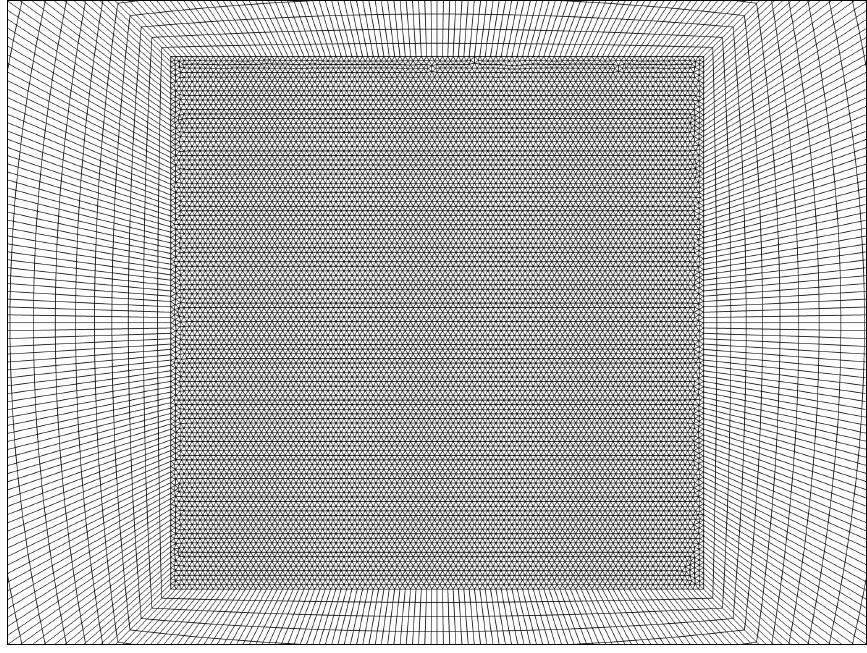}%
  \hfill%
  \includegraphics[width=0.325\linewidth]{\figurespath/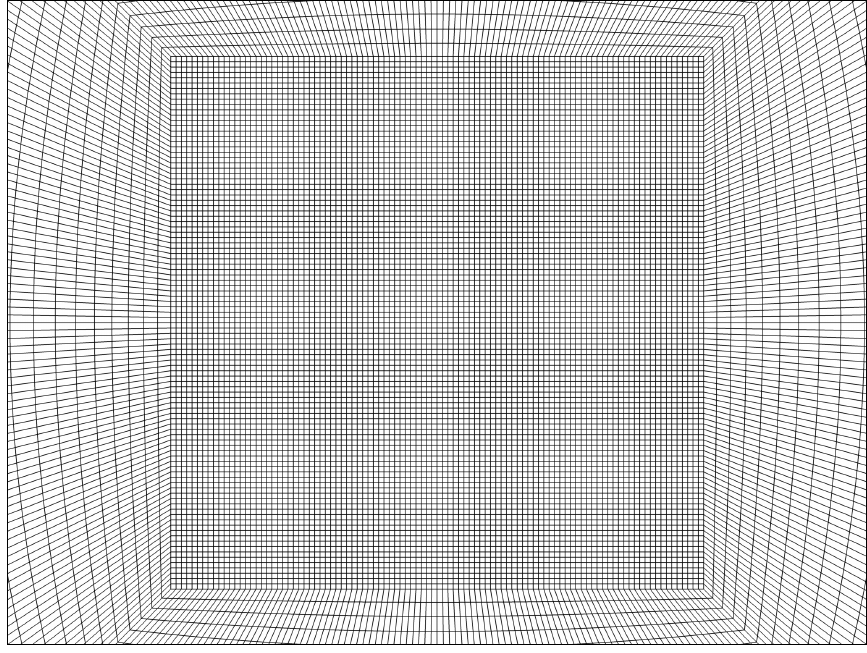}%
  \\[\medskipamount]
  \caption{Initial unstructured background mesh created with GMSH~\cite{geuzaine2009a}, and used in the immersed boundary computations. The full computational domain is a disk of radius $R=500$ (left), with an inner square subdomain with dimensions $[-20,+20]\times[0,1]\times[-20,+20]$. This inner subdomain is made of either hybrid shape elements (center), or only hexahedral elements (right).}%
  \label{fig:numerical-computations:meshes}%
\end{figure*}

All body-fitted meshes have been created with GMSH mesh generator~\cite{geuzaine2009a}. This software was also used to generate the initial background meshes for the immersed boundary computations, which are further refined around the immersed geometry with the preprocessing tool. In \cref{fig:numerical-computations:meshes} are depicted the initial background meshes for the immersed boundary simulations: the full computational domain (left), close-up view of the hybrid unstructured mesh (center) and close-up view of the full hexahedral mesh (right). The initial background mesh is a multiblock circular domain of radius $R=500$. The inner square block has dimensions $[-20,+20]\times[0,1]\times[-20,+20]$.
\subsection{Flow past a cylinder}
We start by considering the flow past a two-dimensional circular cylinder in laminar regime. The governing equations we have used for this test are the Navier--Stokes equations with the immersed boundary volume penalization source terms. We present numerical computations based on a body-fitted mesh and also on hexahedral and hybrid unstructured meshes with immersed geometries.\par
\begin{figure*}
  \centering
  \includegraphics[width=0.325\linewidth]{\figurespath/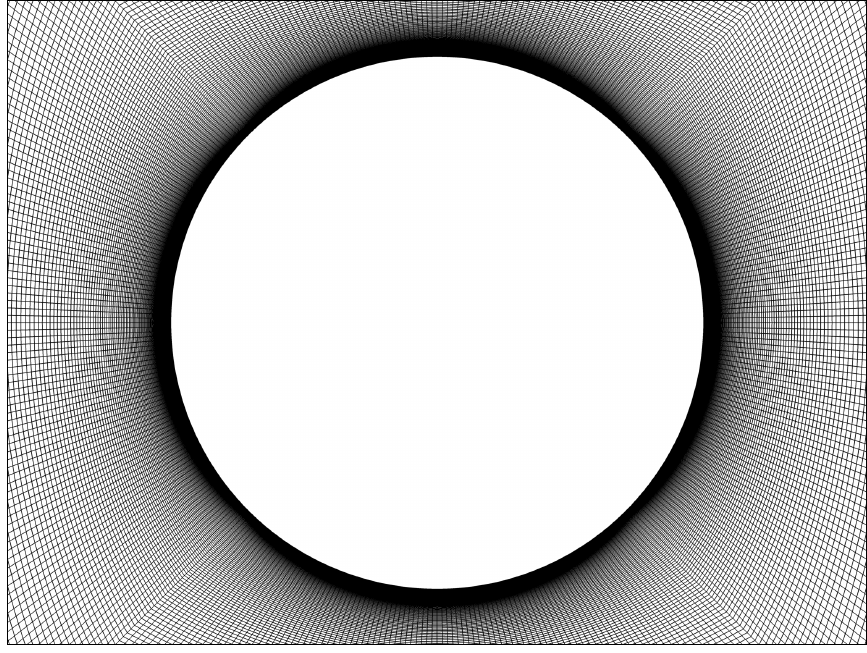}%
  \hfill%
  \includegraphics[width=0.325\linewidth]{\figurespath/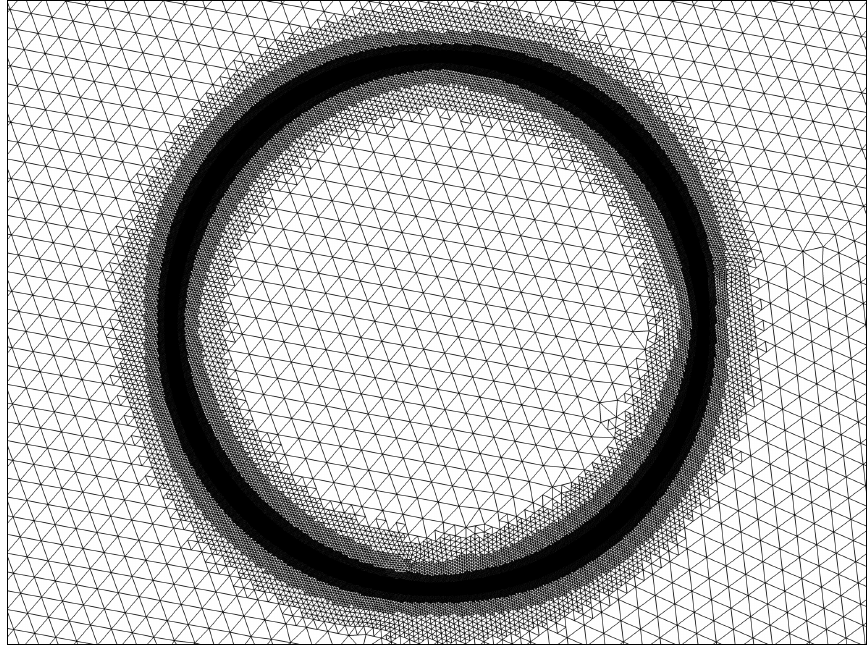}%
  \hfill%
  \includegraphics[width=0.325\linewidth]{\figurespath/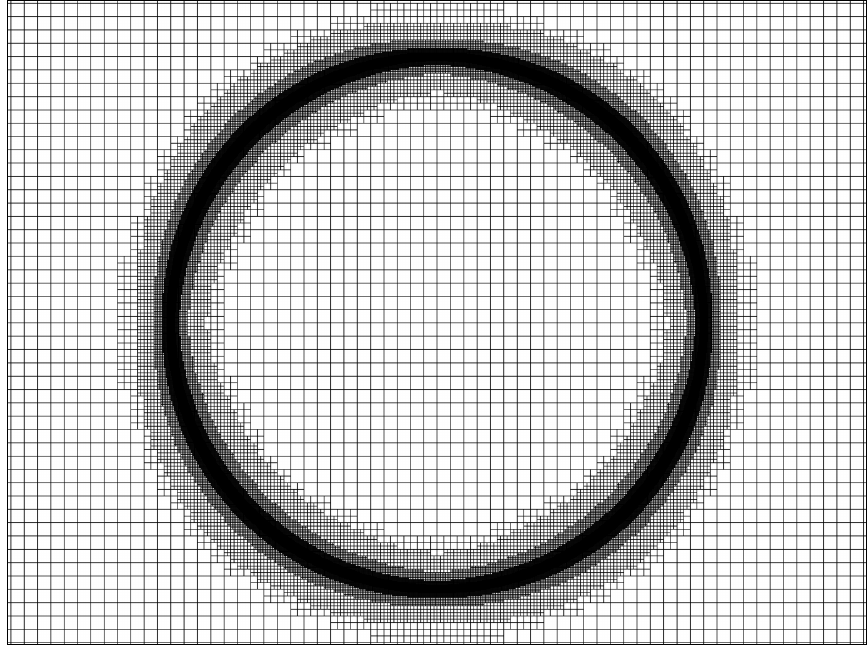}%
  \\[\medskipamount]
  \includegraphics[width=0.325\linewidth]{\figurespath/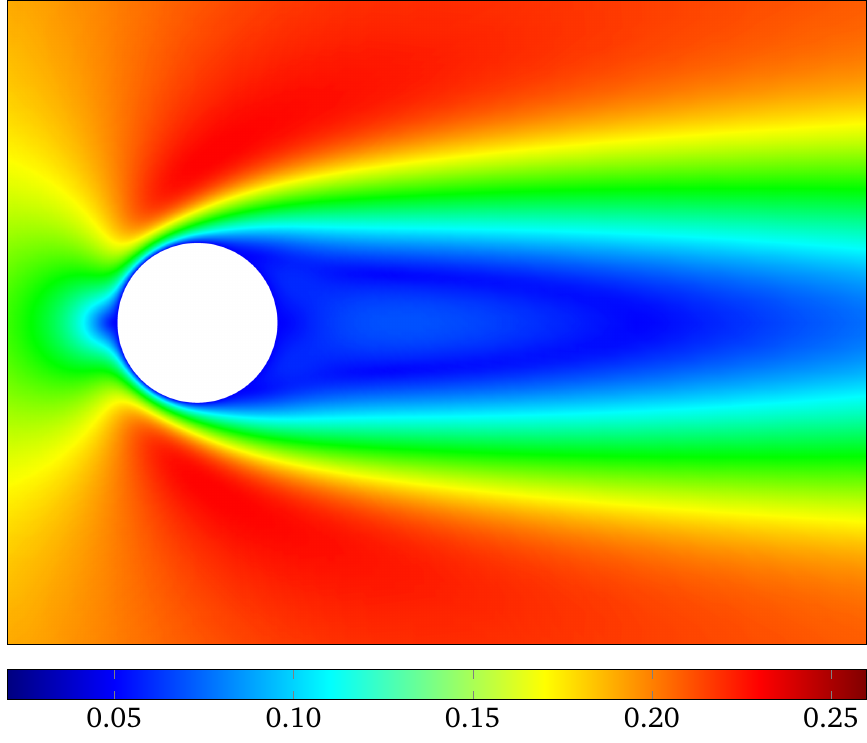}%
  \hfill%
  \includegraphics[width=0.325\linewidth]{\figurespath/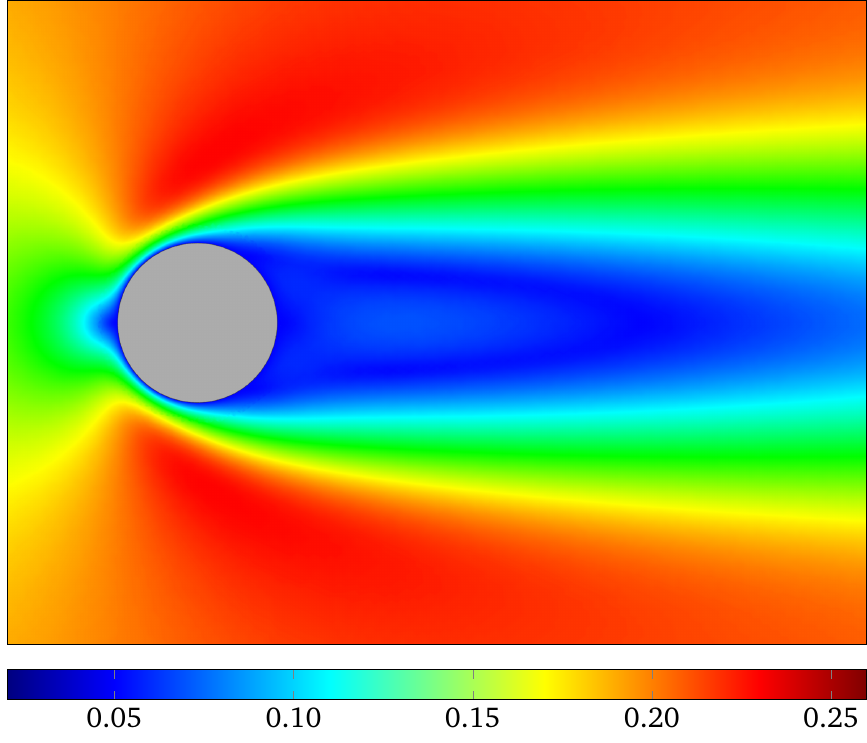}%
  \hfill%
  \includegraphics[width=0.325\linewidth]{\figurespath/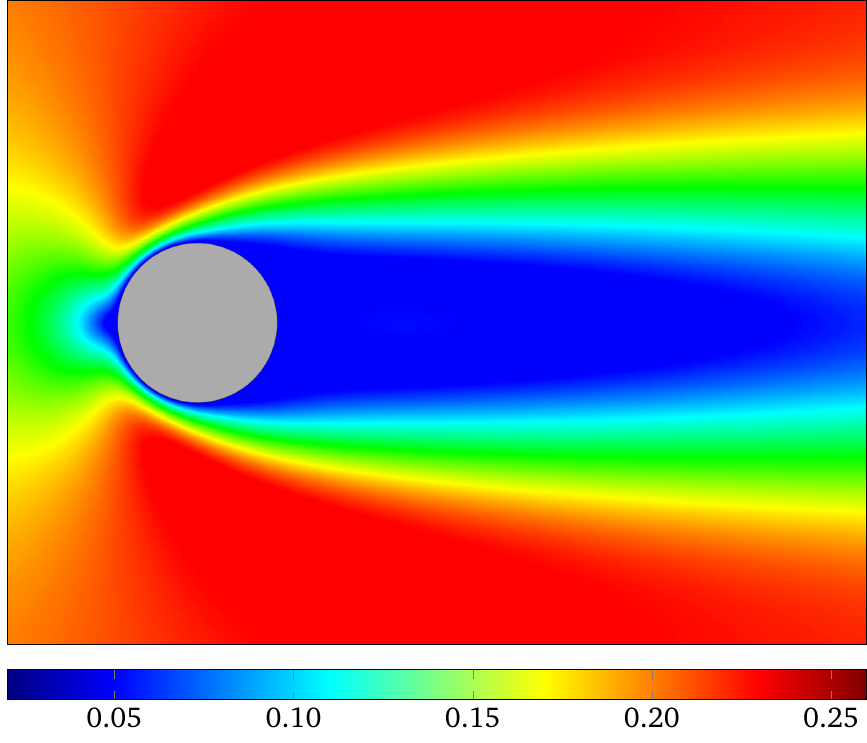}%
  \\[\medskipamount]
  \caption{Flow around a cylinder: computational meshes (top) and contour plots of the velocity magnitude (bottom). The simulation on the body-fitted mesh was performed with a second-order finite volume method (left), the immersed boundary method simulation on the hybrid unstructured mesh with refinement level $l=\num{5}$ was computed with a second-order finite volume method (center), and the immersed boundary method computation on the full hexahedral mesh with refinement level $l=\num{8}$ was computed with a third-order discontinuous Galerkin method (right).}%
  \label{fig:numerical-computations:cylinder:meshes+velocity}%
\end{figure*}

The computational domain is a disk of radius $R=500$ in the $xz$-plane extruded into the $y$-direction a distance $d=\num{1}$. The cylinder has a diameter $D=\num{1}$. We highlight that all meshes have been created with one element in the extrusion direction. The body-fitted mesh is a full hexahedral mesh with elements on the cylinder surface of size $h=\num{5.06E-4}$. This mesh is made of $\num{120000}$ hexahedral elements. Two different background meshes for the immersed boundary method have been created: the first is a hybrid unstructured mesh made of $\num{307637}$ hexahedra and prisms, and the second mesh is a full hexahedral mesh with $\num{169779}$ elements. Both meshes have been refined around the immersed geometry, with a level of refinement $l=\num{5}$ for the hybrid mesh (smallest element size around the immersed geometry $h=\num{1.25E-3}$) and $l=\num{8}$ for the full hexahedral mesh (smallest element size around the immersed geometry $h=\num{1.56E-3}$). The immersed geometry has $\num{440070}$ triangles with characteristic length $\num{5E-3}$. Close-up views of the computational meshes are shown in \cref{fig:numerical-computations:cylinder:meshes+velocity} (top).\par
Regarding flow conditions, the gas has an adiabatic index $\gamma=\num{1.4}$ and flows with Reynolds number $\mathrm{Re}=\num{40}$ and freestream Mach number $M=\num{0.2}$. Under these conditions, the flow remains steady~\cite{canuto2015a}. The non-dimensional density is set to $\rho=\num{1}$, and the non-dimensional pressure $p=\num{1}$. At the boundaries of the computational domain the following boundary conditions were set: far field boundary conditions on the outer face of the disk ($R=500$), and symmetry plane boundary conditions on the front and back faces. For the case of the body-fitted simulation, the cylinder surface was set to the non-slip adiabatic wall boundary condition.\par
The simulation based on a body-fitted mesh was performed with a second-order finite volume method (convergence achieved after $\num{53}$ iterations). The simulation based on the immersed boundary method and using the full hexahedral mesh was calculated with a third-order discontinuous Galerkin method (convergence achieved after $\num{143}$ iterations), and the simulation using the hybrid mesh was calculated with a second-order finite volume method (convergence achieved after $\num{315}$ iterations).
Mesh blanking was not used in immersed-boundary-based simulations, which has an important influence on solver performance~\cite{nunez2025a}. In \cref{fig:numerical-computations:cylinder:meshes+velocity} (bottom) is depicted the contour plot of the velocity magnitude for the three cases considered.\par
\begin{figure*}
  \centering
  \includegraphics[width=1.00\linewidth]{\figurespath/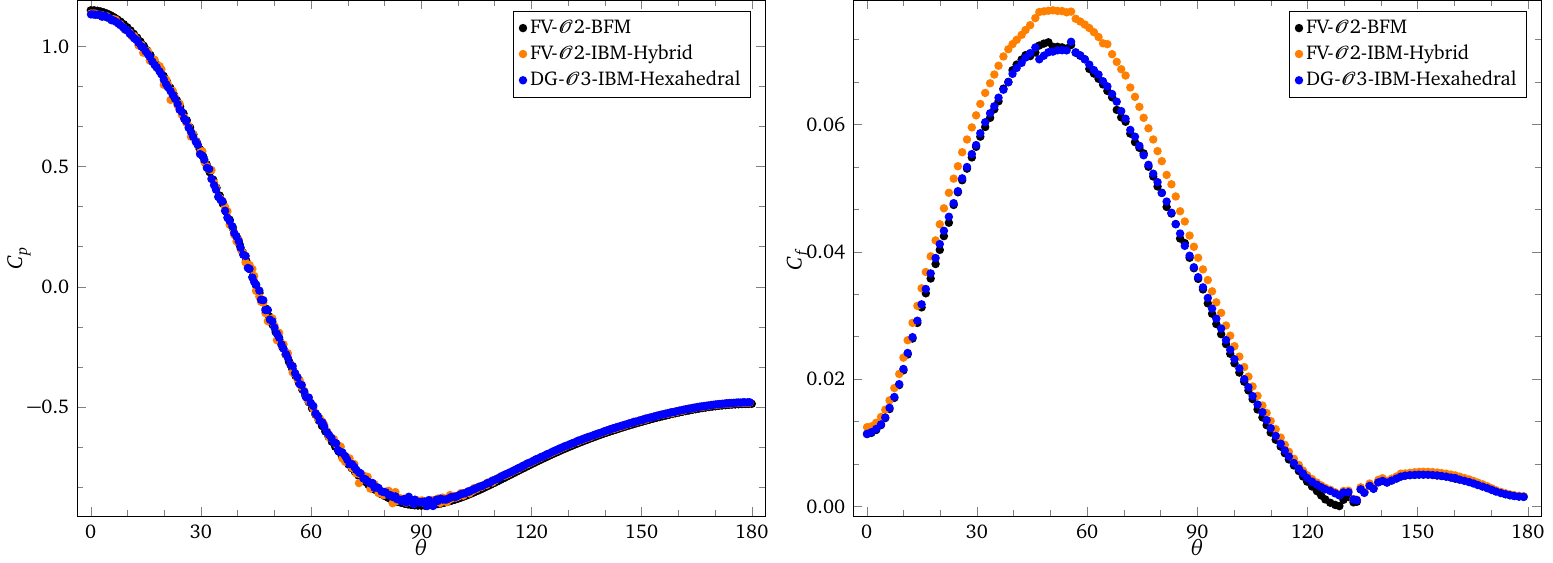}%
  \\[\medskipamount]
  \caption{Flow around a cylinder: comparison of pressure coefficients (left) and skin friction coefficients (right) at $\mathrm{Re}=\num{40}$ and $M=\num{0.2}$. Values were obtained with a second-order finite volume method on a full body-fitted mesh (FV-$\mathcal{O}2$-BFM), a second-order finite volume scheme + immersed boundary method on a hybrid mesh (FV-$\mathcal{O}2$-IBM) and a third-order discontinuous Galerkin method + immersed boundary method on a hexahedral mesh (DG-$\mathcal{O}3$-IBM).}%
  \label{fig:numerical-computations:cylinder:pressure-coefficient+skin-friction-coefficient}%
\end{figure*}

The pressure coefficient $C_{p}$ is plotted in \cref{fig:numerical-computations:cylinder:pressure-coefficient+skin-friction-coefficient} (left). The immersed boundary computations are compared with the full body-fitted simulations based on the same CFD solver. In spite of using lower resolution meshes in the vecinity of the immersed geometry, all results show good agreement on the distribution of the pressure coefficient. Some small oscillations can be observed in the plot associated with the solution computed with the finite volume method on the hybrid mesh. In \cref{fig:numerical-computations:cylinder:pressure-coefficient+skin-friction-coefficient} (right), the skin friction coefficient $C_{f}$ at $y=\num{0.5}$ is depicted. We observe good agreement between the full body-fitted simulation and the solution obtained with the third-order discontinuous Galerkin scheme; the results obtained with the second-order finite volume scheme shows discrepancies. Good agreement for the drag coefficient is shown in \cref{tab:numerical-computations:cylinder:drag-coefficient}. The third-order discontinuous Galerkin solution provides a value of the drag coefficient closer to that obtained with the full body-fitted computation.
\begin{table}
  \centering
  \setlength{\fboxsep}{0.00pt}%
  \renewcommand{\arraystretch}{1.2}%
  \arrayrulecolor{base}
  \newcolumntype{A}{p{0.3\linewidth}}%
  \caption{Flow around a cylinder: drag coefficient at $\mathrm{Re}=\num{40}$ and $M=\num{0.2}$. Values were obtained with a second-order finite volume method on a full body-fitted mesh (FV-$\mathcal{O}2$-BFM), a second-order finite volume scheme + immersed boundary method on a hybrid mesh (FV-$\mathcal{O}2$-IBM) and a third-order discontinuous Galerkin method + immersed boundary method on a hexahedral mesh (DG-$\mathcal{O}3$-IBM).}%
  \label{tab:numerical-computations:cylinder:drag-coefficient}%
  \begin{tabularx}{1.00\linewidth}{XXll}
    \specialrule{2.00pt}{0.00pt}{0.00pt}
    \rowcolor{gray!20}
    Method &
    Mesh &
    $h_{\mathrm{min}}$ &
    $C_{D}$ \\
    \specialrule{2.00pt}{0.00pt}{0.00pt}
    FV-$\mathcal{O}2$-BFM & Full Body-fitted & $\num{5.06E-4}$ & $\num{1.56}$ \\
    FV-$\mathcal{O}2$-IBM & IBM Hybrid       & $\num{1.56E-3}$ & $\num{1.53}$ \\
    DG-$\mathcal{O}3$-IBM & IBM Hexahedral   & $\num{1.25E-3}$ & $\num{1.55}$ \\
    \specialrule{1.00pt}{0.00pt}{0.00pt}
    \multicolumn{3}{l}{\citet{canuto2015a}} & $\num{1.56}$ \\
    \multicolumn{3}{l}{\citet{yu2022a}}     & $\num{1.57}$ \\
    \specialrule{1.00pt}{0.00pt}{0.00pt}
  \end{tabularx}
\end{table}
\subsection{Flow past an NACA0012 airfoil}
The second test we have considered is the flow over an NACA0012 airfoil under weakly compressible ﬂow conditions~\cite{crumpton1993a,jawahar2000a}. The governing equations are again the Navier--Stokes equations. We have performed numerical computations based on a body-fitted mesh and also on hexahedral and hybrid unstructured meshes with immersed geometries. The simulation based on a body-fitted mesh was performed with a second-order finite volume method. The simulation based on the immersed boundary method on the full hexahedral mesh was computed with a second-order discontinuous Galerkin method, and the simulation based on the immersed boundary method on the hybrid mesh was computed with a second-order finite volume method. \par
\begin{figure*}
  \centering
  \includegraphics[width=0.325\linewidth]{\figurespath/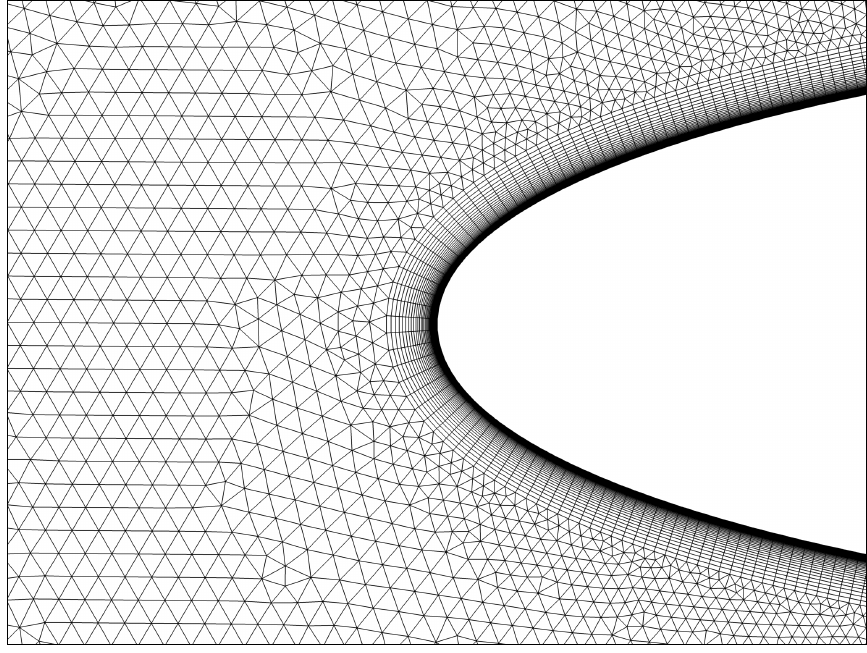}%
  \hfill%
  \includegraphics[width=0.325\linewidth]{\figurespath/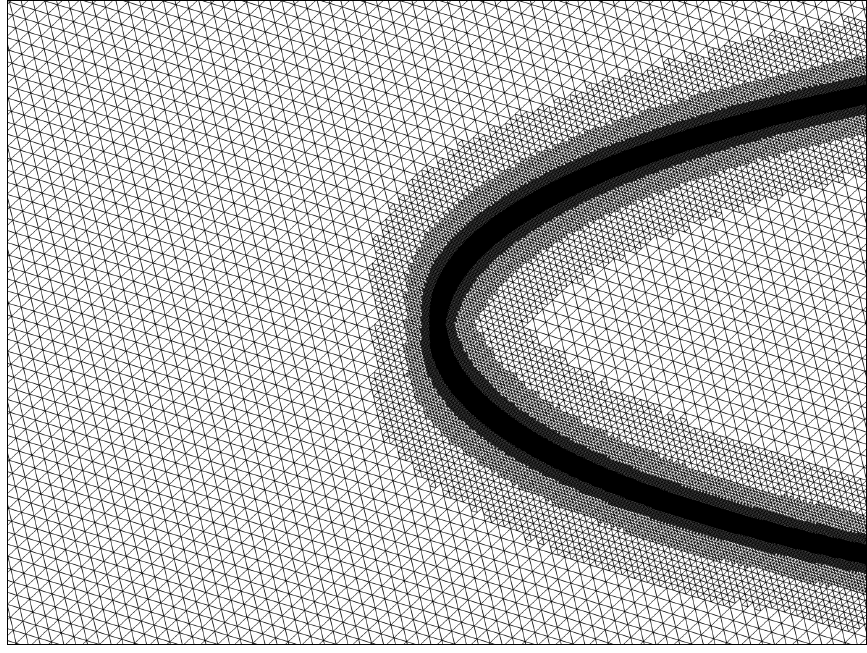}%
  \hfill%
  \includegraphics[width=0.325\linewidth]{\figurespath/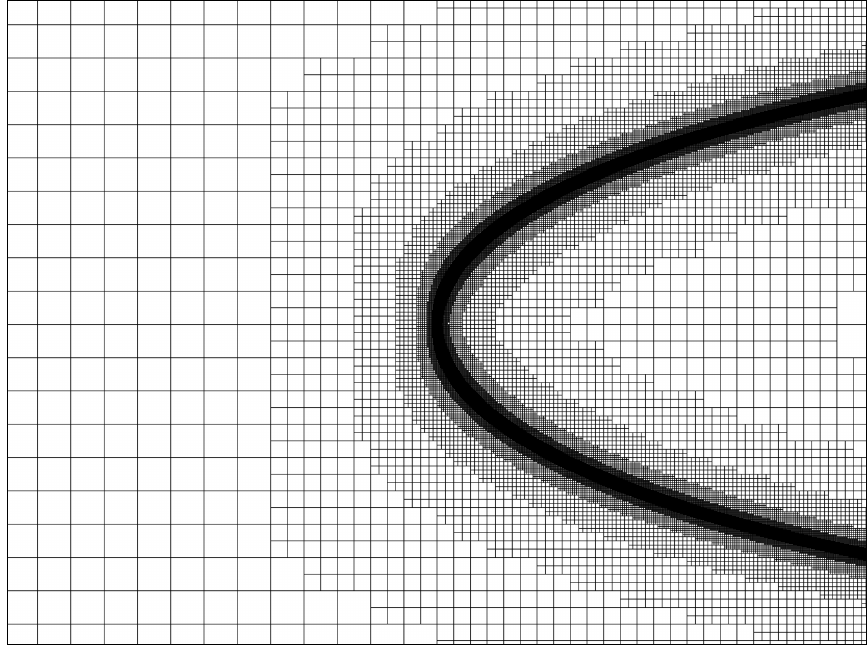}%
  \\[\medskipamount]
  \includegraphics[width=0.325\linewidth]{\figurespath/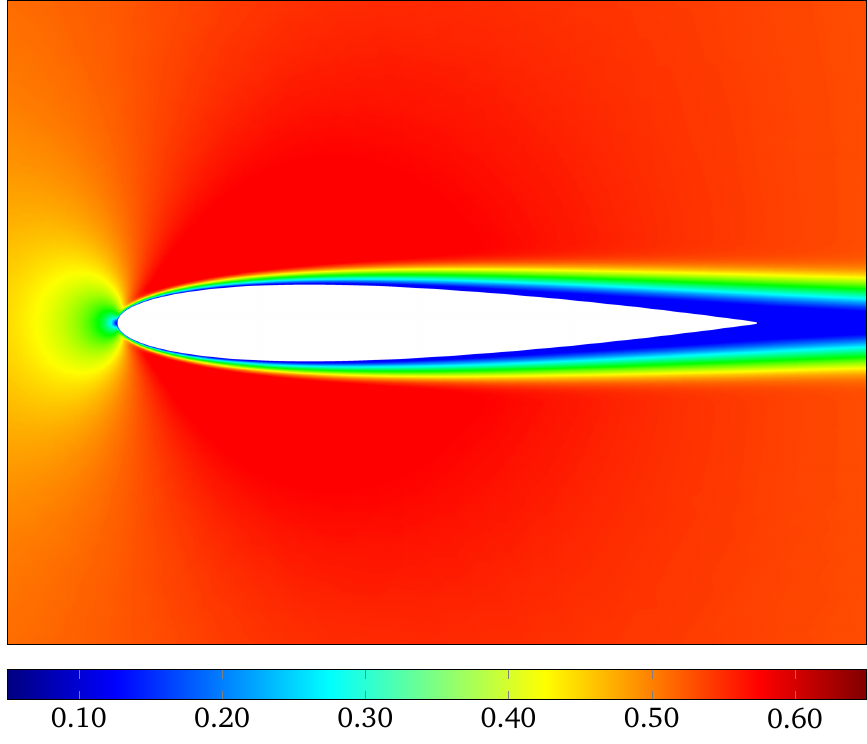}%
  \hfill%
  \includegraphics[width=0.325\linewidth]{\figurespath/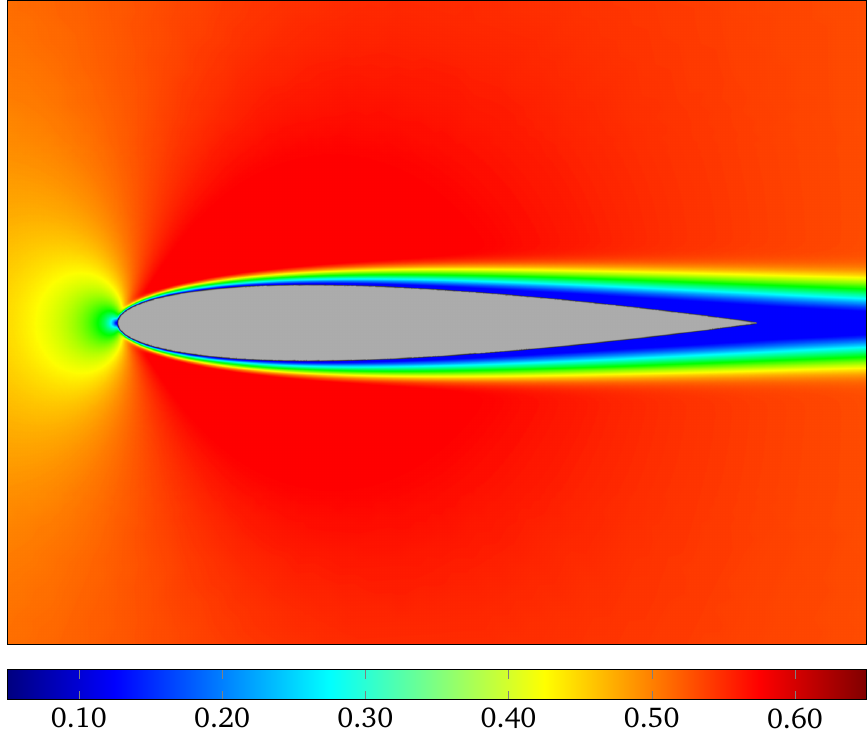}%
  \hfill%
  \includegraphics[width=0.325\linewidth]{\figurespath/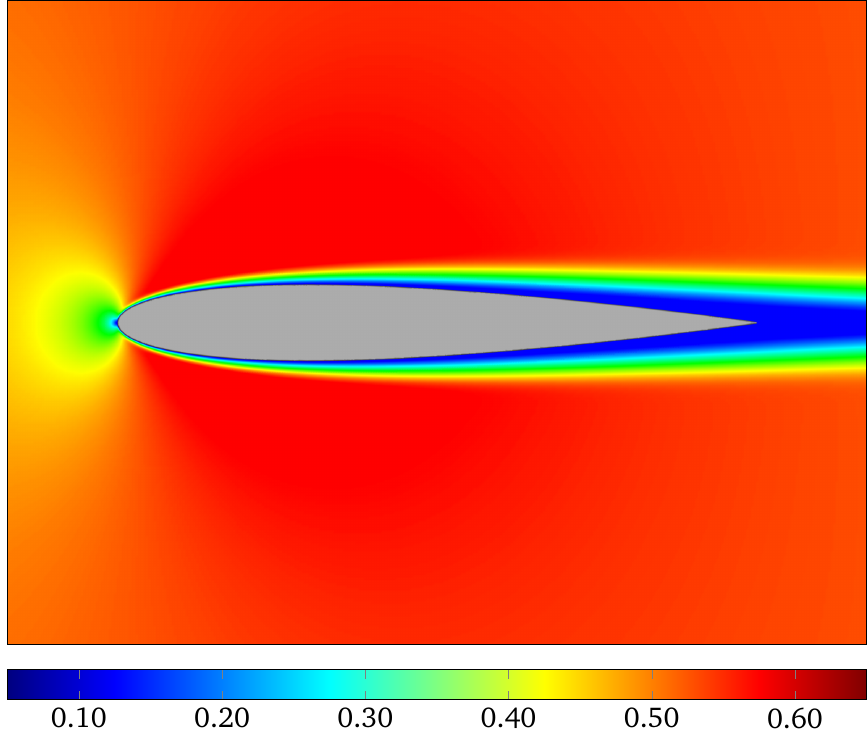}%
  \\[\medskipamount]
  \caption{Flow around an NACA0012 airfoil: computational meshes (top) and contour plots of the velocity magnitude (bottom). The simulation on the body-fitted mesh was performed with a second-order finite volume method (left), the immersed boundary method simulation on the hybrid unstructured mesh with refinement level $l=\num{4}$ was computed with a second-order finite volume method (center), and the immersed boundary method computation on the full hexahedral mesh with refinement level $l=\num{14}$ was computed with a second-order discontinuous Galerkin method (right).}%
  \label{fig:numerical-computations:naca0012:meshes+velocity}%
\end{figure*}

The computational domain is a disk of radius $R=500$ in the $xz$-plane extruded into the $y$-direction a distance $d=\num{1}$. Like in the previous test, all meshes have been created with one element in the extrusion direction. The airfoil chord is $c=\num{1}$. The body-fitted mesh is a hybrid mesh with a layer of hexahedral elements on the surface of the airfoil and a minimum size $h=\num{E-6}$. This mesh is made up of $\num{329731}$ elements. Two different background meshes for the immersed boundary method have been created: the first is a hybrid unstructured mesh made of $\num{1578995}$ hexahedra and prisms, and the second mesh is a full hexahedral mesh with $\num{1867553}$ elements. Both meshes have been refined around the immersed geometry, with a level of refinement $l=\num{4}$ for the hybrid mesh (smallest element size around the immersed geometry $h=\num{1.6E-04}$) and $l=\num{14}$ for the full hexahedral mesh (smallest element size around the immersed geometry $h=\num{6.1E-05}$). The immersed geometry has $\num{817542}$ triangles with characteristic length $\num{2.5E-3}$. The final level of refinement depends on the element size of the initial background mesh in the region to be refined: $h=\num{2.5E-03}$ for the hybrid mesh and $h=\num{3.9E-03}$ for the full hexahedral mesh. Close-up views of the computational meshes are shown in \cref{fig:numerical-computations:naca0012:meshes+velocity} (top).\par
The flow conditions are the following: the gas has an adiabatic index $\gamma=\num{1.4}$ and flows with Reynolds number $\mathrm{Re}=\num{5000}$ and freestream Mach number $M=\num{0.5}$. The simulations were performed with angle of attack $\alpha=\SI{0}{\degree}$. The non-dimensional density is set to $\rho=\num{1}$, and the non-dimensional pressure $p=\num{1}$. At the boundaries of the computational domain the following boundary conditions were set: the outer face of the disk ($R=500$) is set to far field boundary conditions, and the front and back faces are set to symmetry plane boundary conditions. The airfoil surface was set to the non-slip adiabatic wall boundary condition for the case of the body-fitted simulation.\par
The simulation based on a body-fitted mesh was performed with a second-order finite volume method (convergence achieved after $\num{87}$ iterations). The simulation based on the immersed boundary method in the hybrid mesh was calculated with a second-order finite volume method (convergence achieved after $\num{139}$ iterations) and the simulation in the full hexahedral mesh was calculated with a second-order discontinuous Galerkin method (convergence achieved after $\num{147}$ iterations).
The solver stopped iterating once the residual of the conservative variables falls below a predefined threshold ($\delta=\num{E-10}$). In \cref{fig:numerical-computations:naca0012:meshes+velocity} (bottom) is shown the contour plot of the velocity magnitude for the three cases considered.\par
\begin{figure*}
  \centering
  \includegraphics[width=1.00\linewidth]{\figurespath/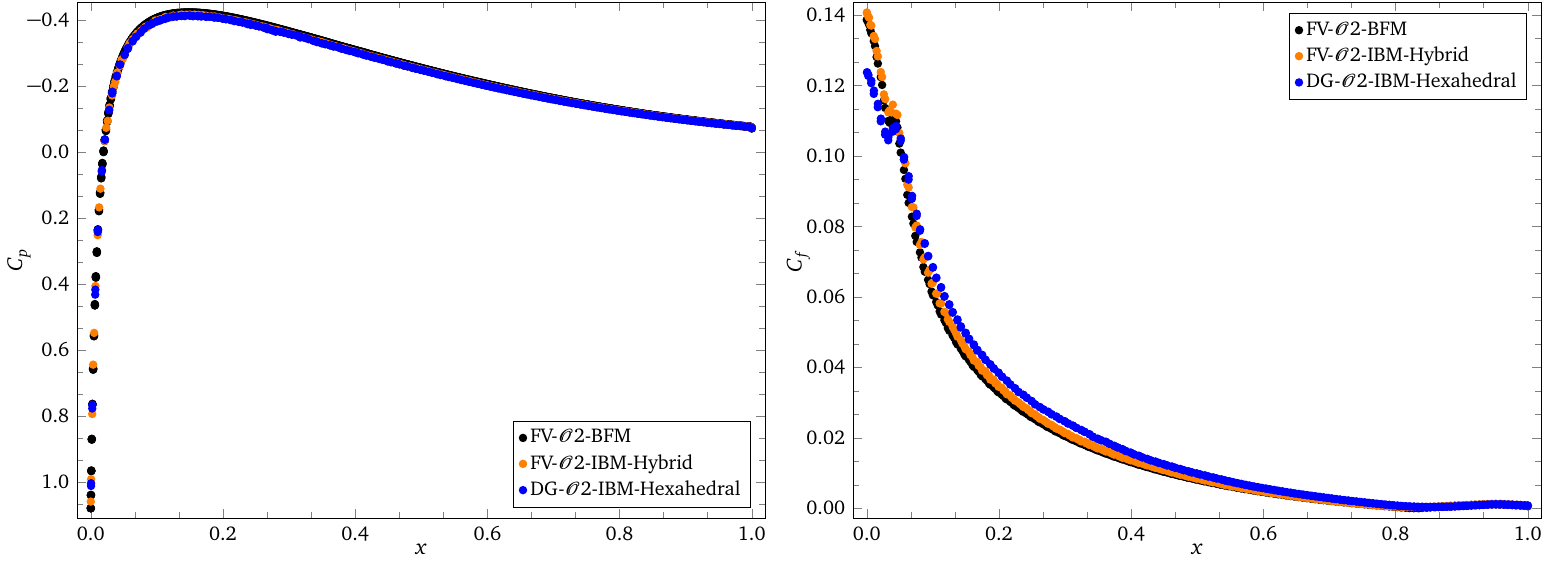}%
  \\[\medskipamount]
  \caption{Flow around an NACA0012 airfoil: comparison of pressure coefficients (left) and skin friction coefficients (right) at $\mathrm{Re}=\num{5000}$, $M=\num{0.5}$ and $\alpha=\SI{0}{\degree}$. Values were obtained with a second-order finite volume method on a full body-fitted mesh (FV-$\mathcal{O}2$-BFM), a second-order finite volume scheme + immersed boundary method on a hybrid mesh (FV-$\mathcal{O}2$-IBM) and a second-order discontinuous Galerkin method + immersed boundary method on a hexahedral mesh (DG-$\mathcal{O}2$-IBM).}%
  \label{fig:numerical-computations:naca0012:pressure-coefficient+skin-friction-coefficient}%
\end{figure*}

In \cref{fig:numerical-computations:naca0012:pressure-coefficient+skin-friction-coefficient} (left), the pressure coefficient $C_{p}$ is plotted for the different meshes and numerical methods used in this test. We observe a good agreement on the distribution of the pressure coefficient. In \cref{fig:numerical-computations:naca0012:pressure-coefficient+skin-friction-coefficient} (right), the skin friction coefficient $C_{f}$ at $y=\num{0.5}$ is plotted. We can observe that the skin friction coefficient exhibits minor local discrepancies when compared to body-fitted reference simulations. This behavior is a known characteristic of immersed boundary formulations due to the stepwise representation of oblique boundaries. As the grid resolution near the immersed boundary is progressively refined, the computed skin friction coefficient curves systematically converge toward the body-fitted reference data. In \cref{tab:numerical-computations:naca0012:drag-coefficient} we show the drag coefficient, which is in good agreement with the values reported in the literature~\cite{crumpton1993a,jawahar2000a,qiu2016a,swanson2016a}.
\begin{table}
  \centering
  \setlength{\fboxsep}{0.00pt}%
  \renewcommand{\arraystretch}{1.2}%
  \arrayrulecolor{base}
  \newcolumntype{A}{p{0.3\linewidth}}%
  \caption{Flow around an NACA0012 airfoil: drag coefficient for computations at $\mathrm{Re}=\num{5000}$, $M=\num{0.5}$ and $\alpha=\SI{0}{\degree}$. Values were obtained with a second-order finite volume method on a full body-fitted mesh (FV-$\mathcal{O}2$-BFM), a second-order finite volume scheme + immersed boundary method on a hybrid mesh (FV-$\mathcal{O}2$-IBM) and a second-order discontinuous Galerkin method + immersed boundary method on a hexahedral mesh (DG-$\mathcal{O}2$-IBM).}%
  \label{tab:numerical-computations:naca0012:drag-coefficient}%
  \begin{tabularx}{1.0\linewidth}{XXll}
    \specialrule{2.00pt}{0.00pt}{0.00pt}
    \rowcolor{gray!20}
    Method &
    Mesh &
    $h_{\mathrm{min}}$ &
    $C_{D}$ \\
    \specialrule{2.00pt}{0.00pt}{0.00pt}
    FV-$\mathcal{O}2$-BFM & Full Body-fitted & $\num{1.0E-06}$  & $\num{0.05576}$ \\
    FV-$\mathcal{O}2$-IBM & IBM Hybrid       & $\num{1.6E-04}$ & $\num{0.05631}$ \\
    DG-$\mathcal{O}2$-IBM & IBM Hexahedral   & $\num{6.1E-05}$ & $\num{0.05592}$ \\
    \specialrule{1.00pt}{0.00pt}{0.00pt}
    \multicolumn{3}{l}{\citet{crumpton1993a}} & $\num{0.05610}$ \\
    \multicolumn{3}{l}{\citet{jawahar2000a}}  & $\num{0.05557}$ \\
    \specialrule{1.00pt}{0.00pt}{0.00pt}
  \end{tabularx}
\end{table}
\subsection{Flow past an MDA30P30N multi-element airfoil}
The third test we have considered in the benchmarking is the flow past a multi-element airfoil, namely the McDonnell Douglas 30P-30N landing configuration (MDA30P30N). Extensive experimental and numerical studies have been conducted for the flow past multi-element airfoils in the last decades~\cite{valarezo1991a,valarezo1992a,rogers1994a,rogers1994b,anderson1995a,klausmeyer1997a,rumsey1998a,spaid2000a}. A very well known fact in the aircraft industry is that the accurate prediction of the flow over multi-element airfoils during high-lift conditions allows to enhance the performance and the safety factor of aircrafts. This airfoil has been extensively tested in the NASA Langley Low Turbulence Pressure Tunnel at various Reynolds and Mach numbers and has also been numerically simulated using a wide range of numerical techniques to solve the Navier--Stokes equations along with different turbulence models.\par
In our simulations, we have solved the Reynolds-averaged Navier--Stokes equations along with the Spalart--Allmaras wall model. This specific wall model significantly improves standard RANS predictions provided that the $y^{+}$ value at the wall-adjacent points remains approximately at or below $\num{60}$, as demonstrated by~\citep{nunez2025a}. The multi-element configuration is characterized by the leading edge slat and the trailing edge flap having a deflection angle of $\SI{30}{\degree}$. Only computations based on a body-fitted mesh for the slat and wing profiles are considered. The immersed geometry corresponds to the flap of the multi-element configuration and, due to the initial background body-fitted mesh, this geometry is placed in a region made of prisms and hexahedra. The equations are solved by the CODA CFD solver based on a second-order finite volume method coupled with the immersed boundary volume penalization method.\par
\begin{figure*}
  \centering
  \includegraphics[width=0.325\linewidth]{\figurespath/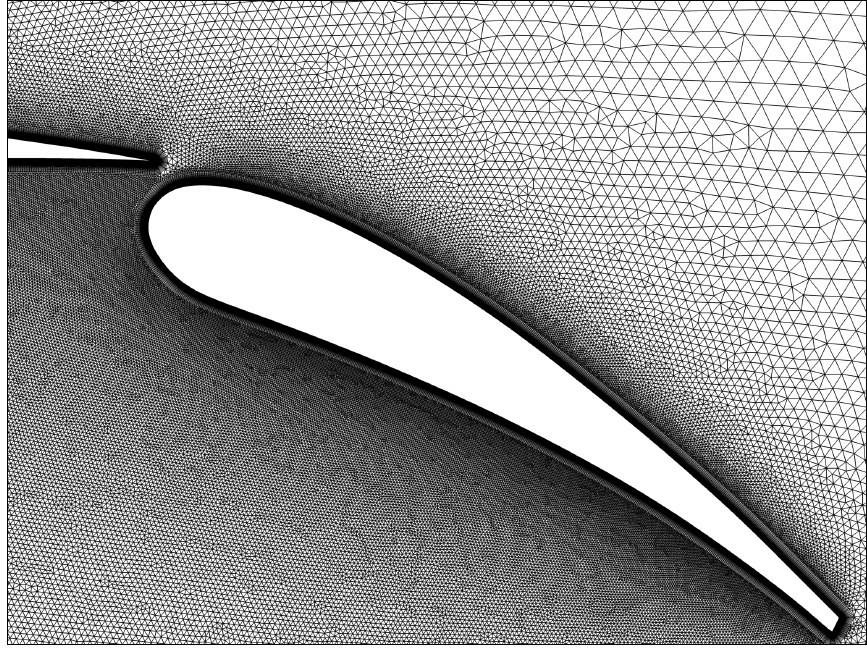}%
  \hfill%
  \includegraphics[width=0.325\linewidth]{\figurespath/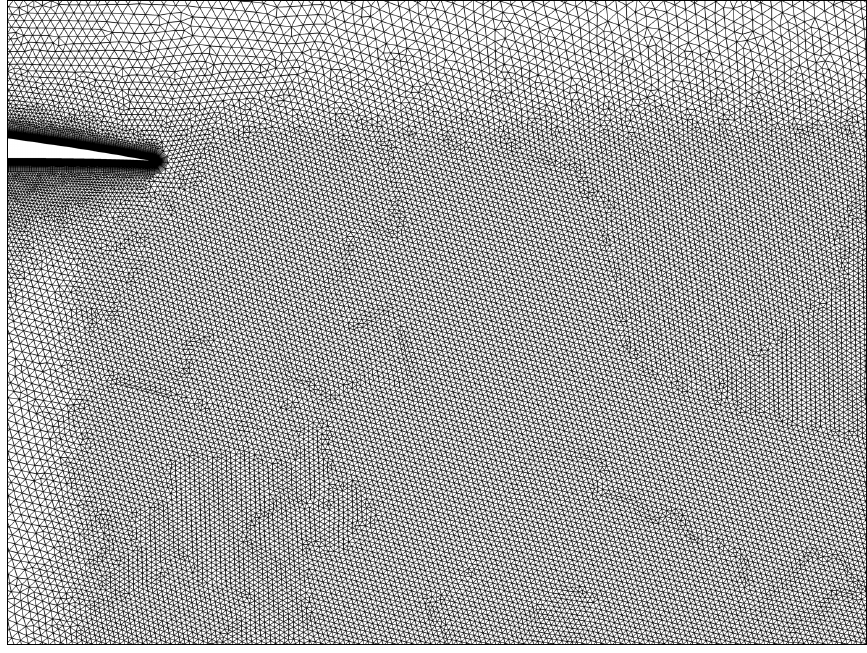}%
  \hfill%
  \includegraphics[width=0.325\linewidth]{\figurespath/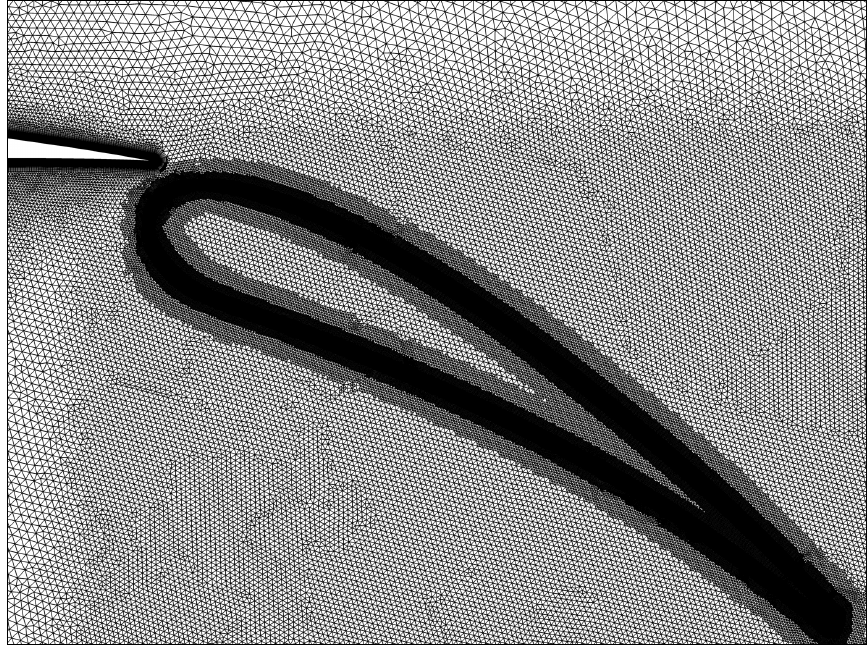}%
  \\[\medskipamount]
  \includegraphics[width=0.325\linewidth]{\figurespath/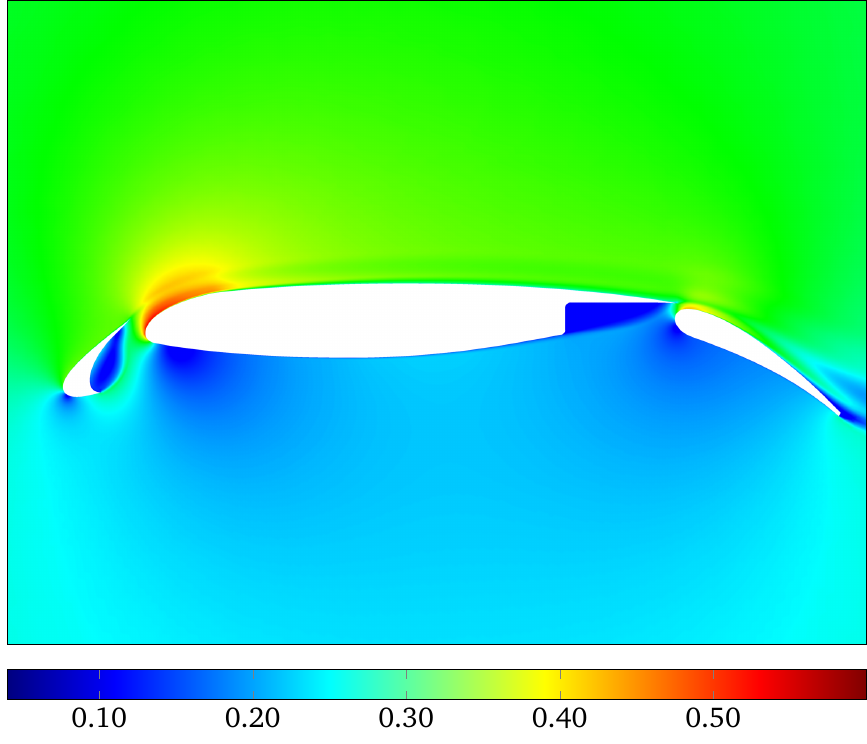}%
  \hfill%
  \includegraphics[width=0.325\linewidth]{\figurespath/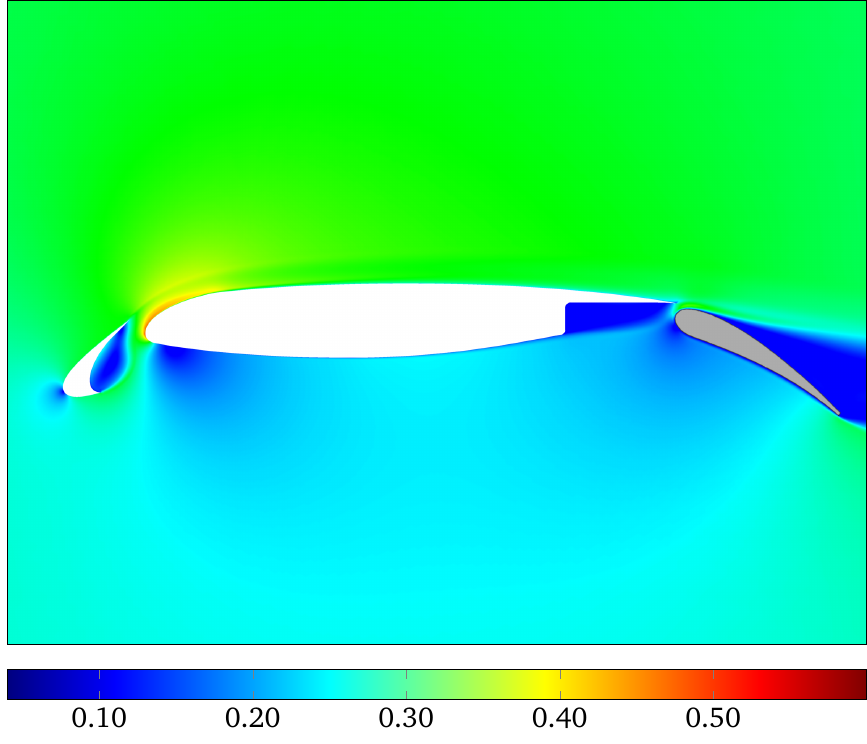}%
  \hfill%
  \includegraphics[width=0.325\linewidth]{\figurespath/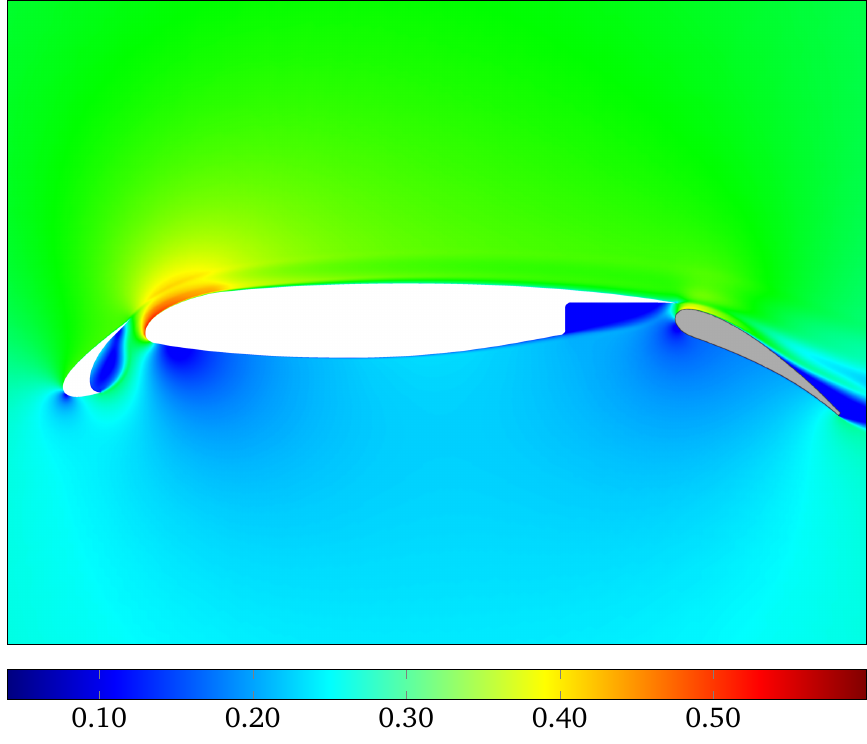}%
  \\[\medskipamount]
  \caption{Flow around an MDA30P30N multi-element airfoil: computational meshes (top) and contour plots of the velocity magnitude (bottom). The simulation on the full body-fitted mesh for the wing-slat-flap system was performed with a second-order finite volume method (left), and a second-order finite volume method coupled with the immersed boundary volume penalization method was used in the simulations on the hybrid unstructured mesh without refinement (center) and on the hybrid mesh with refinement level $l=\num{7}$ (right).}%
  \label{fig:numerical-computations:mda30p30n:meshes+velocity}%
\end{figure*}

Like in the previous tests, the computational domain is a disk of radius $R=500$ in the $xz$-plane, and it has been extruded into the $y$-direction a distance $d=\num{1}$, with one element in the extrusion direction. The full body-fitted mesh is a hybrid mesh with a layer of hexahedral elements on the slat, wing, and flap surfaces. This viscous layer has a starting size from the surface $h=\num{E-6}$. This mesh is made up of $\num{736472}$ elements. We used the computations on this mesh to compare them with those based on immersed boundary methods. Two hybrid meshes for the simulations based on the immersed boundary methods were considered (both meshes have a layer of hexahedral elements only on the slat and wing surfaces, and the flap geometry is not used during the generation of this body-fitted mesh): the first mesh, with $\num{913456}$ elements, has a region where we place the flap, with the smallest element size around the immersed geometry being $h=\num{2.0E-3}$; the second mesh is obtained by refining the previous mesh around the flap surface up to refinement level $l=\num{7}$, getting a mesh with $\num{4585591}$ elements, where the smallest element size around the immersed geometry is $h=\num{1.6E-05}$. The immersed geometry is made of $\num{376196}$ triangles with characteristic length $\num{2.0E-3}$. Close-up views of the computational meshes are shown in \cref{fig:numerical-computations:mda30p30n:meshes+velocity} (top).\par
For this problem, we set the following flow conditions: the gas has an adiabatic index $\gamma=\num{1.4}$ and flows with Reynolds number $\mathrm{Re}=\num{9E6}$ and freestream Mach number $M=\num{0.2}$. For the finest mesh ($h=\num{1.6E-05}$), and these flow conditions, the estimated near-wall grid spacing corresponds to $y^{+}\approx\num{4.8}$. The simulations were performed with angle of attack $\alpha=\SI{5.5}{\degree}$. The non-dimensional density is set to $\rho=\num{1}$, and the non-dimensional pressure $p=\num{1}$. At the boundaries of the computational domain the following boundary conditions were set: the outer face of the disk ($R=500D$) is set to far field boundary conditions, and the front and back faces are set to symmetry plane boundary conditions. At slat and wing surfaces were set with non-slip adiabatic wall boundary conditions. On the boundaries of the computational domain the following boundary conditions were set: far field boundary conditions on the outer face of the disk (at $R=500$), and symmetry plane boundary conditions on the front and back faces. At the surfaces of the slat and wing (and also on the flap for the full body-fitted mesh), non-slip adiabatic wall boundary conditions were established.\par
\begin{figure*}
  \centering
  \includegraphics[width=1.00\linewidth]{\figurespath/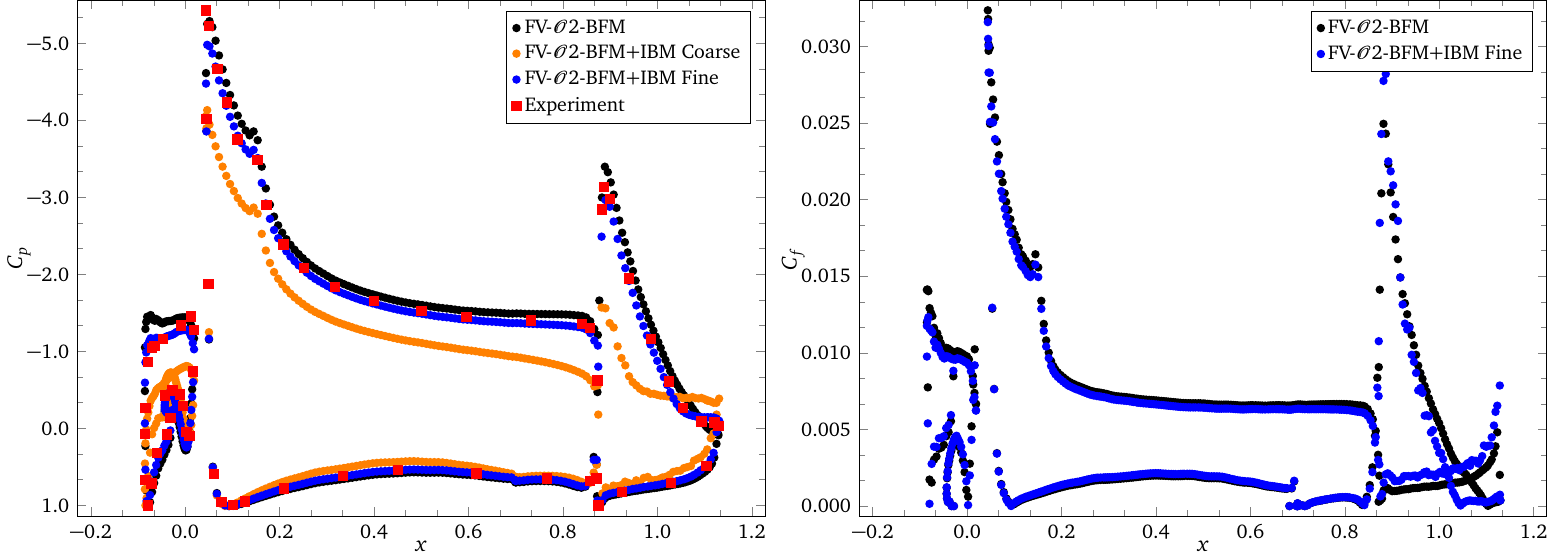}%
  \\[\medskipamount]
  \caption{Flow around an MDA30P30N multi-element airfoil: comparison of pressure coefficients (left) and skin friction coefficients (right) at $\mathrm{Re}=\num{9.0E06}$, $M=\num{0.2}$ and $\alpha=\SI{5.5}{\degree}$. Values were obtained with a second-order finite volume method on a full body-fitted mesh (FV-$\mathcal{O}2$-BFM), a second-order finite volume scheme + immersed boundary method on a coarse hybrid mesh and a fine hybrid mesh (FV-$\mathcal{O}2$-IBM). Experimental data from~\citet{murayama2014b} are also shown. }%
  \label{fig:numerical-computations:mda30p30n:pressure-coefficient+skin-friction-coefficient}%
\end{figure*}

In \cref{fig:numerical-computations:mda30p30n:meshes+velocity} (bottom), the contour plots of the velocity magnitude are depicted for the three cases considered. The solution obtained on the full body-fitted mesh is our reference solution (\cref{fig:numerical-computations:mda30p30n:meshes+velocity} (bottom-left)). In the figure, we can observe a remarkable discrepancy between the flow characteristics on the upper surface of the flap between the simulations on the hybrid unstructured mesh without refinement (bottom-center) and the one on the hybrid mesh with refinement (bottom-right). The computations on the fine mesh are in good agreement with the full body-fitted simulation, except for the small separation region present on the suction side of the upstream side of the trailing edge of the flap. Analysis of the pressure coefficient $C_{p}$ (see \cref{fig:numerical-computations:mda30p30n:pressure-coefficient+skin-friction-coefficient}, left) allows us to confirm the very good agreement between the experimental data (taken from~\citet{murayama2014b}), the full body-fitted mesh calculations and those based on the fine mesh, and also allows us to conclude that the results obtained on the coarse mesh and using immersed boundary methods have a notable disagreement with respect to the experimental data on the upper surface of the slat, wing and flap. In \cref{fig:numerical-computations:mda30p30n:pressure-coefficient+skin-friction-coefficient} (right), the skin friction coefficient $C_{f}$ at $y=\num{0.5}$ is plotted. We observe that the surface skin friction coefficient is very well predicted in the slat and wing parts of the multi-element airfoil, but it exhibits minor local discrepancies when compared to body-fitted reference simulations. This behavior is a known characteristic of immersed boundary formulations due to the stepwise representation of oblique boundaries. As the grid resolution near the immersed boundary is progressively refined, the computed skin friction coefficient curves systematically converge toward the body-fitted reference data. The employment of the immersed boundary methods requires very fine meshes around the immersed geometry. The lift coefficient values are shown in \cref{tab:numerical-computations:mda30p30n:lift-coefficient}. The computations based on immersed boundary methods show a large difference in the lift coefficient for the case of the coarse mesh, but this difference becomes smaller as the mesh is refined, getting closer to the experimental value $C_{L}=\num{2.876}$, taken from~\cite{murayama2014b,wu2023a}.\par
\begin{table}
  \centering
  \setlength{\fboxsep}{0.00pt}%
  \renewcommand{\arraystretch}{1.2}%
  \arrayrulecolor{base}
  \newcolumntype{A}{p{0.3\linewidth}}%
  \caption{Flow around an MDA30P30N multi-element airfoil: lift coefficient at $\mathrm{Re}=\num{9.0E06}$, $M=\num{0.2}$ and $\alpha=\SI{5.5}{\degree}$. Values were obtained with a second-order finite volume method on a full body-fitted mesh (FV-$\mathcal{O}2$-BFM), a second-order finite volume scheme + immersed boundary method on a coarse hybrid mesh and a fine hybrid mesh (FV-$\mathcal{O}2$-IBM). Numerical and experimental data from~\citet{murayama2014b,wu2023a} are also shown.}%
  \label{tab:numerical-computations:mda30p30n:lift-coefficient}%
  \begin{tabularx}{1.0\linewidth}{XXll}
    \specialrule{2.00pt}{0.00pt}{0.00pt}
    \rowcolor{gray!20}
    Method &
    Mesh &
    $h_{\mathrm{min}}$ &
    $C_{L}$ \\
    \specialrule{2.00pt}{0.00pt}{0.00pt}
    FV-$\mathcal{O}2$-BFM & Full Body-fitted & $\num{1.0E-06}$ & $\num{2.847E0}$ \\
    FV-$\mathcal{O}2$-IBM & IBM Hybrid       & $\num{2.0E-03}$ & $\num{2.040E0}$ \\
    FV-$\mathcal{O}2$-IBM & IBM Hybrid       & $\num{1.6E-05}$ & $\num{2.776E0}$ \\
    \specialrule{1.00pt}{0.00pt}{0.00pt}
    \multicolumn{3}{l}{\citet{wu2023a}}       & $\num{2.820E0}$ \\
    \multicolumn{3}{l}{\citet{murayama2014b}} & $\num{2.876E0}$ \\
    \specialrule{1.00pt}{0.00pt}{0.00pt}
  \end{tabularx}
\end{table}
\subsection{Optimization of flap positioning}
Finally, we want to illustrate with the following example an industrial application of the immersed boundary methods along with the mesh refinement tool we have developed. The application consists in optimizing the flap position in a two-element airfoil with respect to the lift coefficient. To this end, we have developed an optimization methodology using the open-source GEMSEO optimization suite and the CODA CFD solver. GEMSEO stands for Generic Engine for Multidisciplinary Scenarios, Exploration and Optimization~\cite{gallard2018a}. GEMSEO is an open source Python software designed to automate multidisciplinary processes, starting with multidisciplinary design optimization ones, and offering a catalog of different algorithms and formulations to make this automation possible.\par
\begin{figure*}
  \centering
  \includegraphics[width=0.49\linewidth]{\figurespath/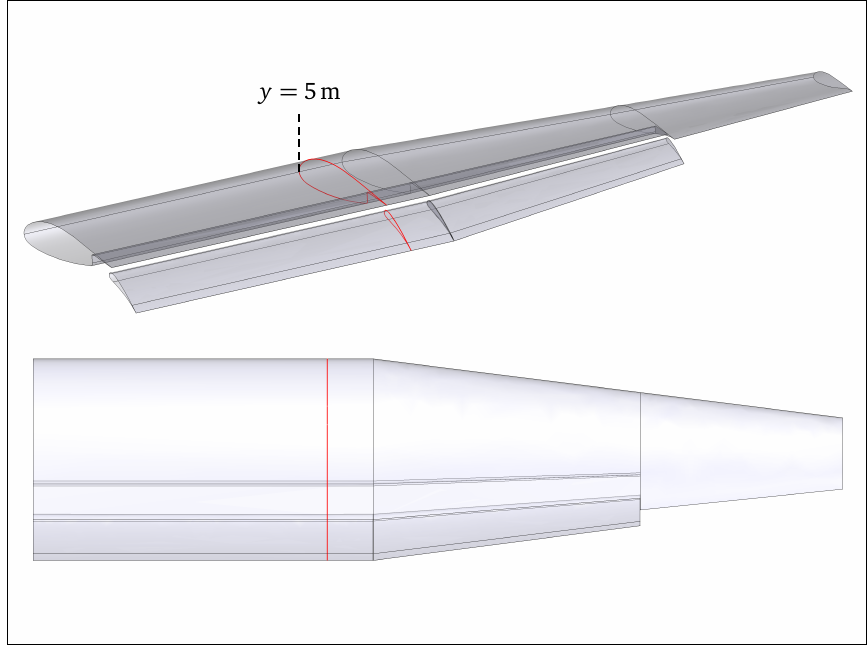}%
  \hfill
  \includegraphics[width=0.49\linewidth]{\figurespath/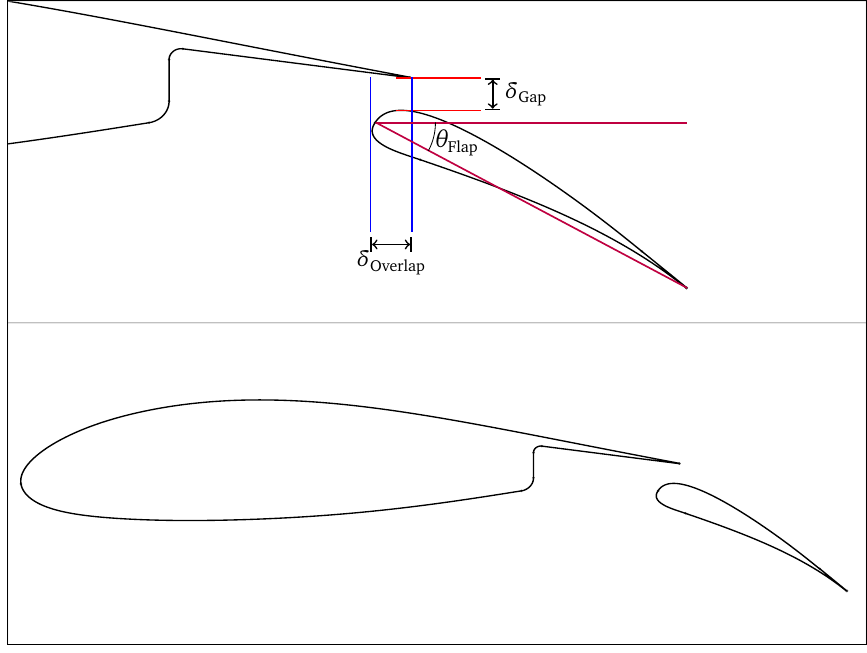}%
  \\[\medskipamount]
  \caption{Wing/flap $\mathrm{3D}$ configuration (left) and its $\mathrm{2D}$ cross-section at $y=\SI{5}{m}$ (right).}%
  \label{fig:numerical-computations:hera-wing-flap:geometry-slice}%
\end{figure*}

Our optimization methodology can use either full body-fitted meshes or meshes with immersed geometries. The first case requires the mesh generation tool to remesh the grid at each optimization function evaluation, while the second case can remesh an initial background mesh via our preprocessing tool to adjust the mesh to the immersed geometry or can simply use an initial background mesh which is already refined in a larger region covering all possible locations of the immersed geometry during the optimization, avoiding the need to remeshing at each function evaluation. Our aim with the optimization procedure is to maximize the lift coefficient under a given high-lift flight condition. We use GEMSEO and CODA CFD solver to perform this task, where the position of the flap is modified and the main airfoil location is held fixed.  We have selected the profile corresponding to the HERA cross-section of the wing/flap at $y=\SI{5}{m}$ from the original $\mathrm{3D}$ setup, as seen in \cref{fig:numerical-computations:hera-wing-flap:geometry-slice} (left). The position of the flap with respect to the wing is characterized by the variables: gap, overlap, and deflection angle. The gap and overlap variables are measured with respect to the trailing edge coordinates of the wing and the leading edge of the flap. The deflection angle is the angle between the chord of the wing and the chord of the flap, as seen in \cref{fig:numerical-computations:hera-wing-flap:geometry-slice} (right).\par
\begin{figure*}
  \centering
  \includegraphics[width=\linewidth]{\figurespath/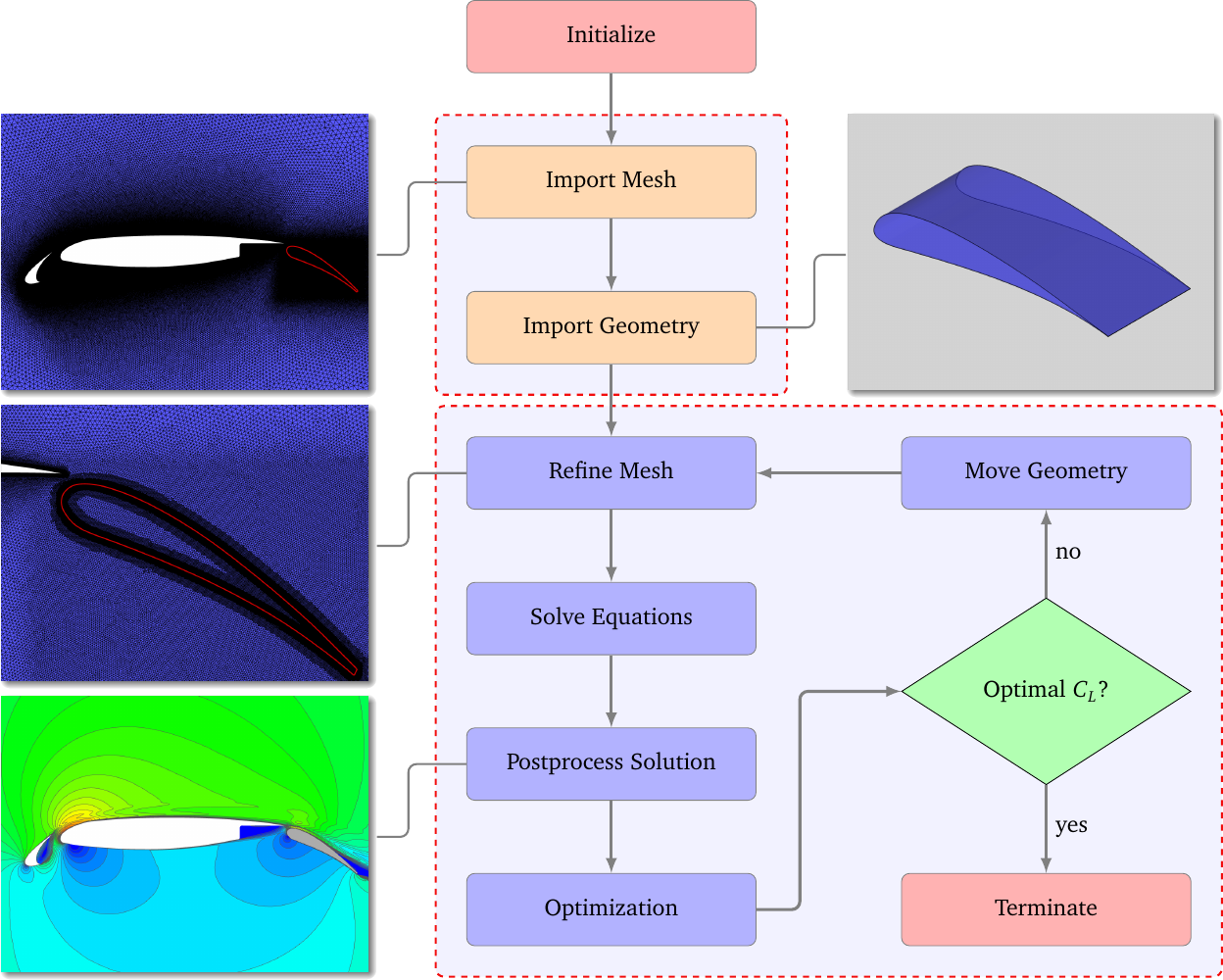}%
  \\[\medskipamount]
  \caption{Flowchart of the optimization methodology used in this work.}%
  \label{fig:numerical-computations:hera-wing-flap:flowchart-optimization}%
\end{figure*}
The optimization is carried out with GEMSEO using the following design variables: angle of attack, gap, overlap, and deflection angle of the flap with respect to the wing. The general workflow is sketched in \cref{fig:numerical-computations:hera-wing-flap:flowchart-optimization}. The objective function (maximum lift or $L/D$) requires calculating the flow through the wing/flap system with the design variables as input variables, and a mesh file created for each configuration of the airfoil. An initial unstructured body-fitted mesh of the domain containing the wing is created, and this mesh is further refined in the region where the flap is located according to the position given by the optimizer. The CODA CFD solver simulates the flow past the wing/flap system for the current position of the flap, with appropriate initial and boundary conditions according to the high-lift flight configuration. In the first optimization step, the output of the objective function is the maximization of the lift coefficient. GEMSEO will call this objective function as many times as the optimization algorithm requires. The COBYQA optimization algorithm is used because of its excellent convergence properties and because it does not require the computation of the adjoint. COBYQA stands for Constrained Optimization BY Quadratic Approximations, and it is designed to supersede COBYLA as a general derivative-free optimization solver. It can handle unconstrained, bound-constrained, linearly constrained, and nonlinearly constrained problems. It uses only function values of the objective and constraint functions, if any. No derivative information is needed. Note that other optimization algorithms are available in GEMSEO and can be adapted to the framework with minimum effort.\par
\begin{figure}
  \centering
  \includegraphics[width=0.75\linewidth]{\figurespath/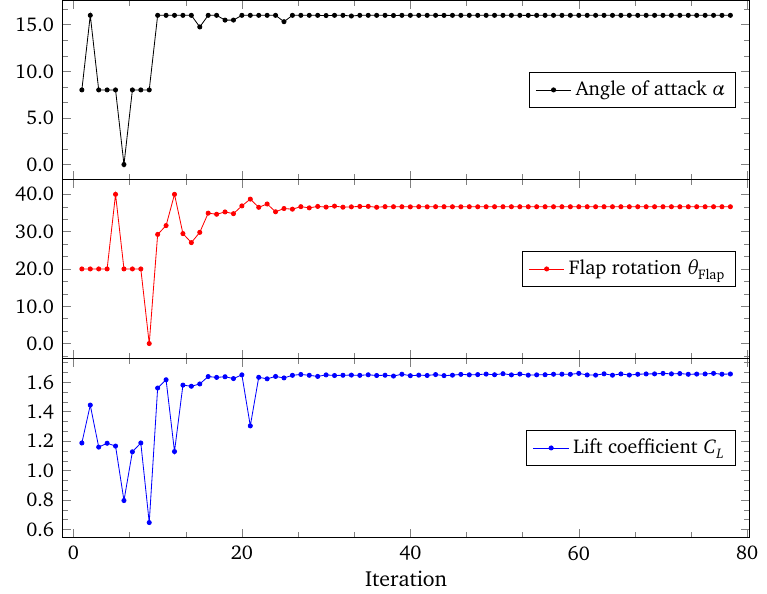}%
  \\[\medskipamount]
  \caption{Optimization of flap position: Convergence history of the optimization process, showing the angle of attack and flap deflection design variables, and the target variable lift coefficient.}%
  \label{fig:numerical-computations:hera-wing-flap:convergence-history}%
\end{figure}
\begin{figure*}
  \centering
  \includegraphics[width=0.49\linewidth]{\figurespath/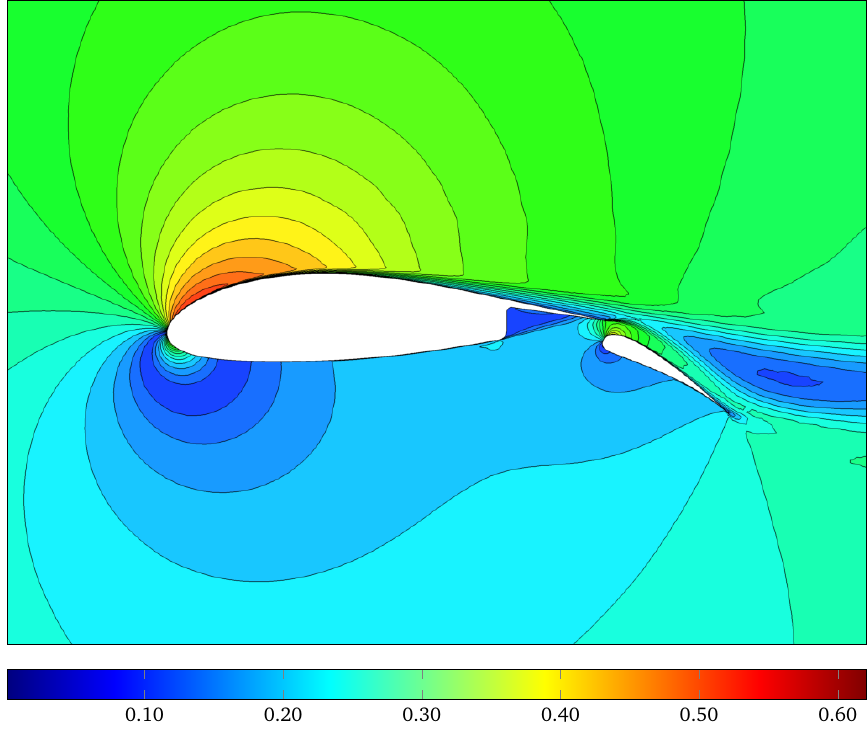}%
  \hfill%
  \includegraphics[width=0.49\linewidth]{\figurespath/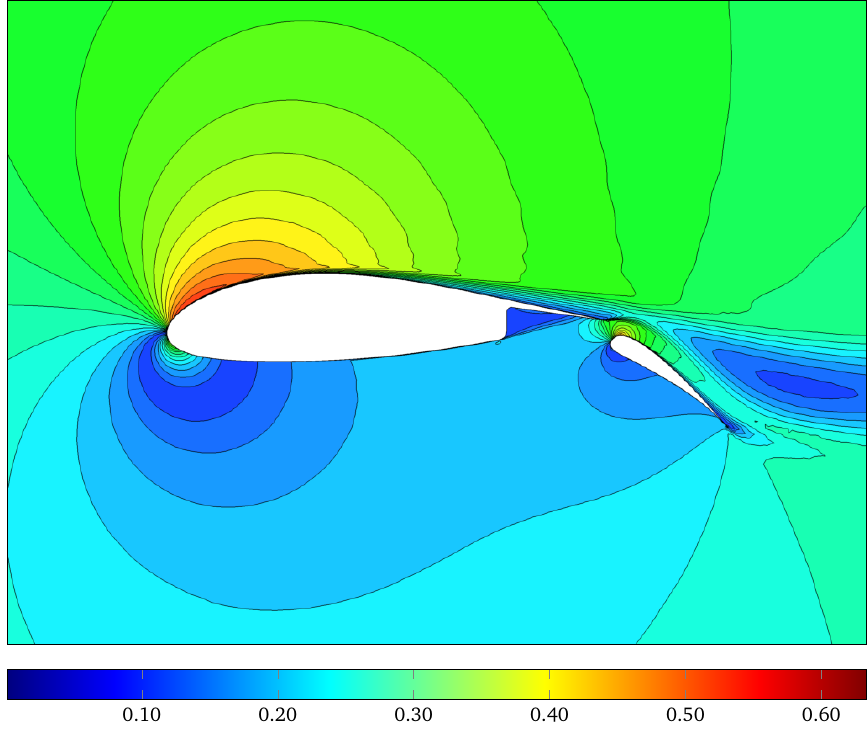}%
  \\[\medskipamount]
  \caption{Optimization of flap position: contour plots of the velocity magnitude for the baseline configuration of the airfoil-flap system (left) and the optimized airfoil-flap configuration (right).}%
  \label{fig:numerical-computations:hera-wing-flap:velocity-magnitude}%
\end{figure*}

The convergence history of the optimization process is presented in \cref{fig:numerical-computations:hera-wing-flap:convergence-history}. We point out that GEMSEO along with the COBYQA algorithm was able to converge after $\num{76}$ iterations, also in $\num{76}$ CODA evaluations. In the figure, the evolution of the lift coefficient for different wing/flap configurations is shown. The values of the design variables for the optimal position of the flap where the lift coefficient is maximum are summarized in \cref{tab:numerical-computations:hera-wing-flap:lift-coefficient}. Additionally, we show the contour plots of the velocity magnitude for the baseline configuration \cref{fig:numerical-computations:hera-wing-flap:velocity-magnitude} (left) and the optimal configuration \cref{fig:numerical-computations:hera-wing-flap:velocity-magnitude} (right).\par
\begin{table}
  \centering
  \setlength{\fboxsep}{0.00pt}%
  \renewcommand{\arraystretch}{1.2}%
  \arrayrulecolor{base}
  \newcolumntype{A}{p{0.3\linewidth}}%
  \caption{Optimization of flap position: design variables values of the original and optimized configurations. The flow conditions are $\mathrm{Re}=\num{12.0E06}$ and $M=\num{0.2}$.}
  \label{tab:numerical-computations:hera-wing-flap:lift-coefficient}%
  \begin{tabularx}{1.0\linewidth}{Xlllll}
    \specialrule{2.00pt}{0.00pt}{0.00pt}
    \rowcolor{gray!20}
    Configuration               &
    $\alpha$                    &
    $\delta_{\mathrm{Overlap}}$ &
    $\delta_{\mathrm{Gap}}$     &
    $\theta_{\mathrm{Flap}}$    &
    $C_{L\mathrm{max}}$        \\
    \specialrule{2.00pt}{0.00pt}{0.00pt}
    Original  & $\SI{16}{\degree}$ & $\num{0.0}$   & $\num{0.0}$       & $\SI{30}{\degree}$    & $\num{1.5880}$ \\
    Optimized & $\SI{16}{\degree}$ & $\num{0.053}$ & $\num{0.000252}$ & $\SI{36.68}{\degree}$ & $\num{1.6557}$ \\
    \specialrule{1.00pt}{0.00pt}{0.00pt}
  \end{tabularx}
\end{table}
%
\section{Conclusions}\label{sec:conclusions}
We have developed a new preprocessing tool that enables geometry-adaptive mesh refinement on hybrid unstructured meshes for immersed boundary method simulations, providing flexibility for complex geometries in computational fluid dynamics. The preprocessing tool places immersed bodies in unstructured body-fitted meshes made of different types of elements, such as tetrahedra, hexahedra, prisms, and pyramids, and can refine the mesh around the surface of the geometry to enhance accuracy and efficiency in complex computational fluid dynamics simulations. This approach can be used to treat only geometrical details (control surfaces, ice shapes, etc.) as immersed in a body-fitted, wall-resolved mesh of a clean aerodynamic surface. The unstructured hybrid meshes refined with our preprocessing tool are compatible with the finite volume and discontinuous Galerkin methods coupled with the immersed boundary volume penalization scheme.\par
The main data structures and algorithms used in the development of the mesh preprocessing tool are explained in detail. Arrays, linked lists, hash tables and $k$-d trees data structures were implemented to allow the processing and refinement of dynamic meshes with different types of elements and with hanging nodes. The mesh is stored in a linked list, and the immersed geometry in an array. Hash tables and $k$-d trees were used to distribute the geometry in the mesh and to create the interfaces that establish the relationship of the master face nodes with the slave face nodes. Efficient algorithms to compute the overlap of the mesh elements with the facets of the geometry were taken from the current computer graphics state-of-the-art algorithms.\par
Several numerical computations were performed by solving the Navier--Stokes and Reynolds-Averaged Navier--Stokes equations: the subsonic flow past a cylinder, the subsonic flow past an NACA0012 airfoil, and the subsonic flow past an MDA30P30N multi-element airfoil. As input meshes, full hexahedral and hybrid unstructured meshes were used for the simulations based on finite volume and discontinuous Galerkin methods coupled with the immersed boundary volume penalization scheme. In spite of using lower resolution meshes in the vicinity of the immersed geometry, all results are in good agreement with the corresponding full body-fitted computations and also with the experimental data for the pressure coefficient curves and the lift and drag coefficients. Our findings allow us to conclude that finer meshes are required if an excellent agreement is sought, but, as a consequence, the number of elements of the refined mesh will increase. Convergence can be accelerated if mesh blanking is used. In addition to the cases considered on the test bench, an industrial application example was investigated: optimization of the flap position in a two-element airfoil with respect to the lift coefficient. An optimization methodology was developed using the open-source GEMSEO optimization framework, the CODA CFD solver, and our mesh preprocessing tool. GEMSEO used the COBYQA optimization algorithm, and the optimization was performed using maximum lift as the objective function. The results were consistent with a similar optimization workflow with only body-fitted meshes generated by GMSH.\par
Future work will include the extension of the mesh preprocessing tool to parallel systems and with acceleration performed with graphics processor units. A parallel version of the code will enable the generation of refined meshes with up to a billion elements in a flexible and efficient manner. These large meshes are typical in complex three-dimensional geometries, such as an aircraft or a wind turbine with moving blades.\par
%
\printcredits
%
\section*{Acknowledgments}
Jonatan Núñez-de la Rosa, Esteban Ferrer, and Eusebio Valero acknowledge the funding received by the Grant NextSim\slash AEI\slash 10.13039\slash 501100011033 and H2020, Grant Agreement No. 956104. This project has received funding from the Clean Aviation Joint Undertaking under the European Union's Horizon Europe Research and Innovation Programme under Grant Agreement HERA (Hybrid-Electric Regional Architecture) No. 101102007. However, the views and opinions expressed are those of the author(s) only and do not necessarily reflect those of the European Union or CAJU. Neither the European Union nor the granting authority can be held responsible for them. Finally, all authors gratefully acknowledge the Universidad Politécnica de Madrid (\url{www.upm.es}) for providing computing resources on the Magerit Supercomputer.
%
\bibliographystyle{cas-model2-names}
\bibliography{references}
%
\end{document}